\documentclass[10pt,final,journal,twocolumn]{IEEEtran}
\usepackage{graphicx}
\usepackage{subfigure}
\usepackage[cmex10]{amsmath}
\usepackage{amsthm}
\usepackage{amssymb}
\usepackage{bm}
\usepackage{algorithm}
\usepackage{algorithmic}
\usepackage{cite}
\usepackage{setspace}
\usepackage{stfloats}
\usepackage{enumerate}
\usepackage{cases}
\usepackage{multirow}
\usepackage{mathrsfs}
\usepackage{array}
\usepackage{color}
\usepackage{verbatim}
\usepackage{extpfeil}
\usepackage{hyperref}
\usepackage{booktabs}
\usepackage{multirow}
\usepackage{caption}
\usepackage{graphicx}
\usepackage{tabularx}
\usepackage{epstopdf}
\usepackage{textcomp}
\usepackage{epsfig,epsf,color,balance,cite}
\usepackage{amssymb}
\usepackage{amsthm}
\usepackage{graphicx}
\usepackage{array,color}
\usepackage{amsmath}
\usepackage{graphicx}
\usepackage{tabularx}
\usepackage{epsfig,epsf,color,balance,cite}
\usepackage{algorithmic}
\usepackage{algorithm}
\usepackage{bm}
\usepackage{caption}
\usepackage{textcomp}
\usepackage{color}
\usepackage{multirow}
\usepackage{cite}
\usepackage{enumerate}
\usepackage{cases}
\usepackage{color}
\usepackage{url}
\usepackage{epstopdf}
\usepackage{textcomp}
\def\BibTeX{{\rm B\kern-.05em{\sc i\kern-.025em b}\kern-.08em
    T\kern-.1667em\lower.7ex\hbox{E}\kern-.125emX}}
\begin{document}

	\makeatletter
	\newcommand{\rmnum}[1]{\romannumeral #1}
	\newcommand{\Rmnum}[1]{\expandafter \@slowromancap \romannumeral #1@}
	\makeatother
	
	\title{RIS-Aided Localization Algorithm and Analysis: Tackling Non-Gaussian Angle Estimation Errors}
	
	\author{Tuo Wu, Hong Ren, Cunhua Pan, Yijin Pan, Sheng Hong,  Maged Elkashlan, \\ Feng Shu and Jiangzhou Wang, \emph{Fellow, IEEE}
		
		\thanks{(Corresponding author: \emph{Hong Ren and Cunhua Pan}).
			
			T. Wu and M. Elkashlan are with the School of Electronic Engineering and Computer Science at Queen
			Mary University of London, London E1 4NS, U.K. (Email:\{tuo.wu, maged.elkashlan\}@qmul.ac.uk). H. Ren, C. Pan and  Y. Pan  are with the National Mobile Communications Research Laboratory, Southeast University, Nanjing 210096, China. (Email: \{cpan, hren, panyj\}@seu.edu.cn). S. Hong is with Information Engineering School of Nanchang University, Nanchang 330031, China. (Email: shenghong@ncu.edu.cn).
			F. Shu is with the School of Electronic and Optical Engineering, Nanjing
			University of Science and Technology, Nanjing 210094, China, and also
			with the School of Information and Communication Engineering, Hainan
			University, Haikou 570228, China.(E-mail: shufeng0101@163.com). J.
			Wang is with the School of Engineering, University of Kent, UK. (e-mail: J.Z.Wang@kent.ac.uk). }
		
	}
	
	\markboth{}
	{}
	
	\maketitle

	\begin{abstract}
Reconfigurable intelligent surface (RIS)-aided localization systems are increasingly recognized for enhancing accuracy in internet of things (IoT) networks. However, prevailing studies  tend to either assume a Gaussian distribution for angle estimation errors (AEE) or directly neglect the impact of the AEE, overlooking its non-Gaussian nature in real-world scenarios, particularly with diverse estimation methods (e.g., 2D-DFT algorithm). Addressing this oversight, this paper explores the design and performance analysis of RIS-aided localization systems, specifically tackling non-Gaussian AEE. We adopt the classical two-step three-dimensional (3D) localization scheme to determine the position of mobile user (MU). Initially, we estimate angles of arrival (AoAs) and time differences of arrival (TDoAs) at the RIS using different methods, resulting in non-Gaussian and Gaussian errors, respectively. Subsequently,  to accommodate the non-Gaussian nature of  AoAs errors and the Gaussian character of  TDoA errors, we design a multiple weighted least squares (mWLS) algorithm to accurately localize MU.  Besides, our research also includes a unique bias analysis for evaluating the performance of the proposed localization algorithm under both Gaussian and non-Gaussian errors. Simulation results demonstrate the effectiveness of both the proposed mWLS algorithm and the bias analysis methodology.\end{abstract}

	\IEEEpeerreviewmaketitle

\section{Introduction}
{ As the sixth generation (6G) Internet of Things (IoT) wireless networks continue to develop, there's a growing focus on improving their ability to precisely determine locations \cite{Renzo2}, e.g., smart factories \cite{Ren4}, autonomous vessels \cite{Dai5}, automated vehicles \cite{Renzo5}, and mobile user (MU) sensing \cite{Renzo3}.  To support these applications, a lot of research and development is being conducted in the area of wireless localization algorithms specifically for 6G IoT networks.}

{ Traditional IoT wireless localization algorithms utilize  a two-step localization scheme for localization.   At the first step, channel parameters, e.g., angles of arrival (AoAs) and time differences of arrival (TDoAs), are estimated using channel estimation methods \cite{Dardari, QianTD, Gao1}.    At the second step, non-linear equations that represent the geometric relationships between these channel parameters and the position coordinates are formulated and solved.}

{ However, traditional IoT wireless localization systems typically utilize numbers of  base stations (BS) as reference nodes to achieve high accuracy, leading to increased deployment and hardware costs. Conversely, the reconfigurable intelligent surface (RIS) \cite{Cpan, Panyj1, Ren1, Lee4} not only offers a more cost-effective solution for localization, but it also brings other advantages \cite{Renzo1, Pan2, Zhoug1, Zhou2, Zhi}.  First, RIS can establish line-of-sight (LoS) communication when BS-MU links are blocked \cite{Zhi1}. Second, the slim and  adaptable design of the RIS allows it easy to be integrated on the IoT urban environment \cite{Panyj1, Daill3}. Third, as a passive element, RIS achieves accurate estimations with low energy consumption \cite{Gao1}, due to its large amount of reflecting elements.
	
The above-mentioned advantages of the RIS-aided localization have sparked considerable research, especially in the areas of performance analysis and localization algorithm design. For the area of performance analysis, to investigate the theoretical bound of localization accuracy, Cramer-Rao lower bound (CRLB) and root mean square error (RMSE) have been extensively studied in \cite{JHe} and \cite{Liuy}. Building upon this, Elzanaty \emph{et al.} in \cite{Alouini} examined the impact of multiple RISs in mmWave systems and the effects of signal synchronization on localization accuracy. For the area of localization algorithm design, various algorithms have been developed to tackle RIS-aided localization algorithm design's challenges, ranging from optimizing the  RIS phase shifts \cite{WangR, ZFeng, HZhang, JHe1} to employing RSS-based methods \cite{HZhang} and maximum likelihood (ML) estimation techniques \cite{Wym}.}

Research in RIS-aided localization systems has significantly advanced. However, the impact of angle estimation error (AEE) on AoA based localization algorithms is frequently overlooked. Specifically, many studies fail to incorporate AEE into their algorithm designs, while performance analyses often assume AEE follows a Gaussian distribution. In practice,  the distribution of the AEE depends on the practical estimation methods (e.g., 2D-DFT algorithm \cite{Zhou1}), which may not follow the Gaussian distribution. Therefore, designing algorithms that effectively address non-Gaussian AEE is crucial for accurately reflecting real-world conditions. Analyzing algorithms that consider non-Gaussian AEE characteristics is also valuable, as it can inform network managers about deploying IoT wireless positioning systems more effectively.

Besides, most of the existing researches investigating RIS-aided localization mainly based on only one channel parameter, such as AoAs. However, designing the corresponding localization algorithm based on AoA alone will actually waste the good localization performance of RIS \cite{Wym}, and the corresponding localization accuracy will not be high enough \cite{Cpan}. Therefore, we consider integrating AoA and TDoA to design the localization algorithm.

However, according to the classical localization literatures like \cite{2012Bias}, TDoA errors often follow a Gaussian distribution due to the nature of time measurement techniques, which typically involve averaging multiple signal observations to reduce randomness, leading to a central limit theorem effect. Hence, it remains another non-trivial challenge to design the localization algorithm. To be specific, incorporating non-Gaussian AEE and Gaussian TDoA estimation error  into algorithm design significantly complicates the process due to complex mathematical transformations and heightened optimization requirements. Furthermore, it is another non-trivial difficulty to analyze the localization performance and analyze the performance. To be specific, traditionally deriving the  CRLB in the context of non-Gaussian AEE can be complex, making it difficult to  gain insights through comparisons between the CRLB and the RMSE.

The weighted least square (WLS) method \cite{2002A} is an effective choice for handling both Gaussian and non-Gaussian errors simultaneously due to its ability to assign different weights to different estimation, thus accurately accommodating the varying nature and complexities of these error distributions.  However, it often suffers from limitations such as sensitivity to initial value selection and potential convergence to local minima.  Therefore, it is still challengeable to design an advanced  WLS based algorithm that both utilizes the  AoA and the TDoA, harnessing their different error properties to enhance localization  accuracy.

For an effective performance analysis of the proposed algorithm, we adopt bias as a new evaluation metric in this paper. This choice is driven by the limitations of CRLB in scenarios involving non-Gaussian AoA and Gaussian TDoA errors. The CRLB assumes a likelihood function that is often ill-suited for non-Gaussian distributions, leading to inaccurate performance bounds in such cases. Moreover, the sophisticated mathematical operations required for CRLB, like the computation and inversion of the Fisher Information, are less practical for our analysis. In contrast, bias analysis focuses on calculating the expectation of the estimator and comparing it with the true parameter value, thereby simplifying the evaluation process. It also offers greater flexibility in handling various distribution types and nonlinear effects, making it a more intuitive and adaptable method for evaluating estimator performance. However, analyzing localization performance through bias analysis remains challenging due to the inherently complex nature of WLS, which often involves non-linear relationships and varying error variances across data points. This complexity introduces difficulties in accurately modeling and quantifying the bias, especially since traditional linear assumptions are not applicable in such scenarios.

To address the above-mentioned challenges, we design a multiple WLS (mWLS) localization algorithm to enhance localization  accuracy, aiming at tackling the non-Gaussian AEE. This algorithm jointly utilizes the AoAs with their associated Non-Gaussian estimation errors and the TDoAs with Gaussian estimation errors, enabling the derivation of the closed-form expression for the 3D position of the MU. Additionally, we decompose the expression of the expectation of the error to simplify the calculation and conduct the bias analysis to assess the performance of the proposed system. Our contributions can be summarized as:

	\begin{itemize}
		\item[1)] We employ a two-step localization scheme for IoT mmWave systems aided by multiple RISs. In the initial step, we estimate the AoAs and TDoA, highlighting that the AoA errors exhibit non-Gaussian properties while TDoA errors are Gaussian. We also define the geometric relationship of the MU's position. Then, our uniquely designed mWLS algorithm leverages both types of errors, capitalizing on their distinct properties, to accurately determine the MU's position based on the estimated parameters and their distinct covariances.
		
		\item[2)] To rigorously evaluate our proposed localization algorithm, we undertake a thorough analysis to deduce the theoretical estimation bias, uniquely factoring in both Gaussian and non-Gaussian errors. By decomposing matrices that encompass AoA and TDoA estimation into true values and estimation errors, we ascertain the preliminary estimation error and its theoretical bias. Using this information, we further determine the refining estimation's bias. This nuanced approach, more streamlined than traditional CRLB, clearly highlights the accuracy of our localization algorithm.
		
		\item[3)] Simulation results are provided to evaluate the effectiveness of the proposed localization algorithm for non-Gaussian AEE. Besides, these results not only demonstrate the superior performance of the proposed algorithm, but also validate the applicability of the employed bias analysis for non-Gaussian AEE. Finally, the derived bias results are in good agreement with the simulation results, indicating that the performance of the proposed algorithm in the real-world applications aligns with theoretical expectations.
	\end{itemize}

	The remainder of the paper is organized as follows. The system model for the two-step localization aided by multiple RISs is described in Section \ref{System Model}. The framework of two-step localization scheme is given in Section \ref{frame}.  Section \ref{biasa} derives the bias analysis of the proposed localization scheme. Simulation results are given in Section \ref{simulation}. Section \ref{Con} concludes this work.
	\section{System Model} \label{System Model}
	Consider an  RIS-aided localization system, where an MU sends pilot signals to the BS to locate the MU with the assistance of multiple RISs. RISs could support high data rate while maintaining low costs and energy consumption. Besides, RISs can constructively reflect the signal from the BS to users.  The BS is equipped with a uniform linear array (ULA) of $N_b$ antennas, and the MU is equipped with a single antenna.  Moreover, there are $M$ RISs, and each RIS is a uniform planar array (UPA) \footnote{ {  By installing GPS receivers on devices such as BS and RIS, microsecond time accuracy can be achieved, thus ensuring network-wide synchronization.}}.
	
	The BS is placed parallel to the x-axis with the center located at $\mathbf{p}=[x_p,y_p,z_p]^T$. The $i$-th RIS is placed parallel to the y-o-z plane with its center located at $\mathbf{s}_i = [x_i,y_i,z_i]^T$, $i = 1,2,\cdots,M$. The UPA-based RIS has $N_{y,z}=N_y\times N_z$ reflecting elements, where $N_y$ and $N_z$ denote the numbers of reflecting elements along the y-axis and z-axis, respectively. The true position of the MU is $\mathbf{q}=[x_q,y_q,z_q]^T$ and is assumed to be placed parallel to the x-o-y plane. The estimated location of the MU is $\mathbf{\hat{q}}=[\hat{x}_q,\hat{y}_q,\hat{z}_q]^T$. Generally, once the RISs and BS have been deployed, the coordinates $\mathbf{s}_i$ and $\mathbf{p}$ are known and invariant.  In order to locate the MU, we need to obtain the estimated $\mathbf{\hat{q}}$.
\subsection{Wireless Channel Model}
{ In this subsection, we present the channel model of the system under consideration.} It is important to note that the model includes two types of communication links characterized by distinct channel parameters: the direct link from the MU  to the BS, and the reflecting links via the RISs.

For the direct link from the MU to the BS, we assume that the number of propagation paths between the  BS and the MU is $M$, the azimuth AoA of the $m$-th path from the MU to the BS is $0\leq\theta_{Ua,m}\leq\pi$.  The array response vector at the BS of the $m$-th path can be expressed as
 \begin{align}\label{v31}
{\bf a}_{Ua}(\theta_{Ba,m})=[1,\cdots,e^{\frac{-j2\pi(N_{b}-1) d_b\sin\theta_{Ua,m}}{\lambda_c}}]^T,
\end{align}
where $d_b$ and $\lambda_c$ denote the distance between the antennas of the BS and  the carrier wavelength, respectively. The channel response of the  direct link is given by
\begin{align}\label{v32}
{\bf h}_{BU}=\sum^{M}_{m=1}\delta_{m}{\bf a}_{Ua}(\theta_{Ua,m}),
\end{align}
where $\delta_{m}$ denotes the complex channel gain of the $m$-th path.

{  For the reflecting links between the MU and the BS via RISs, we decompose them into two sub-links: the MU-RISs link and the RISs-BS link.

For the MU-RISs link,} it is assumed that the number of propagation paths between the  MU and the $i$-th RIS is $N_{i}$, the AoA of the $n_i$-th path from the MU to the $i$-th RIS can be decomposed into the elevation angle $0\leq\theta_{Ra,n_i}\leq\pi$ in the vertical direction, and the azimuth angle $0\leq\phi_{Ra,n_i}\leq\pi$ in the horizontal direction, respectively.  The array response vector at the $i$-th RIS of the $n_i$-th path is written as
\begin{align}\label{v33}
{\bf a}_{Ra}(\theta_{Ra,n_i},\phi_{Ra,n_i})={\bf a}_{Ra}^{(e)}(\phi_{Ra,n_i})\otimes{\bf a}_{Ra}^{(a)}(\theta_{Ra,n_i},\phi_{Ra,n_i}),
\end{align}
where $\otimes$ denotes the Kronecker product. Moreover, we have
\begin{align}\label{v34}
&{\bf a}_{Ra}^{(e)}(\phi_{Ra,n_i})=[1,\cdots,e^{\frac{-j2\pi(M_{x}-1) d_r\sin\phi_{Ra,n_i}}{\lambda_c}}]^T,\nonumber\\
&{\bf a}_{Ra}^{(e)}(\theta_{Ra,n_i},\phi_{Ra,n_i})=[1,\cdots,e^{\frac{-j2\pi(M_{z}-1) d_r\cos\theta_{Ra,n_i}\cos\phi_{Ra,n_i}}{\lambda_c}}]^T,
\end{align}
where $d_r$ denotes the distance between the elements of the RISs. The channel of MU-RISs link  can be modeled as
\begin{align}\label{v35}
{\bf h}_i &=\sum^{N_i}_{n_i=1}\alpha_{n_i}{\bf a}_{Ra}(\theta_{Ra,n_i},\phi_{Ra,n_i})\nonumber\\
&= \underbrace{\alpha_{MR,i}{\bf a}_{Ra}(\theta_{MR,i},\phi_{MR,i})}_{LoS}\nonumber\\
&\quad+ \underbrace{\sum^{N_i}_{n_i=2}\alpha_{n_i}{\bf a}_{Ra}(\theta_{Ra,n_i},\phi_{Ra,n_i})}_{NLoS},
\end{align}
where  $\alpha_{n_i}$ denotes the complex channel gain of $n_i$-th path. Moreover, $\alpha_{MR,i}$, $\theta_{MR,i}$, and $\phi_{MR,i}$ denote the complex channel gain, the elevation AoA, and the azimuth AoA of the line-of-sight (LoS) path, respectively. As we can see from \eqref{v35}, channel components of ${\bf h}_i$  can be categorized into two types, namely \emph{LoS} and \emph{NLoS}. LoS path component is the direct path between the BS and the MU, non-line-of-sight (NLoS) path component consists of the paths between the RISs and the MU reflected by scatters, e.g., walls, human bodies.

{ For the RISs-BS link,} it is assumed that the number of propagation paths between the BS and the $i$-th RIS is $J_i$. The  angle of departure (AoD) of the $j_i$-th path from the $i$-th RIS to the BS can be decomposed into the azimuth angle $0\leq\theta_{Rd,j_i}\leq\pi$ in the horizontal direction and the elevation angle $0\leq\phi_{Rd,j_i}\leq\pi$ in the vertical direction, respectively. The array response vector is given by ${\bf a}_{Rd}(\theta_{Rd,j_i},\phi_{Rd,j_i})$, which is similar to the expression of ${\bf a}_{Ra}(\theta_{Ra,n_i},\phi_{Ra,n_i})$ in \eqref{v33}.

Besides, let us define the azimuth AoA  at the BS as $0\leq\theta_{RB,j_i}\leq\pi$, the array response vector ${\bf a}_{RB}(\theta_{RB,j_i})$ can be written as
 \begin{align}\label{v36}
{\bf a}_{RB}(\theta_{RB,j_i}) =[1, \cdots,e^{\frac{-j2\pi(N_{b}-1) d_r\sin\theta_{RB,j_i}}{\lambda_c}}]^T.
\end{align}
 By using the array response vector ${\bf a}_{Rd}(\theta_{Rd,j_i},\phi_{Rd,j_i})$ and ${\bf a}_{RB}(\theta_{RB,j_i})$, the channel matrix of the $i$-th RIS with the BS can be formulated as
	\begin{align}\label{v37}
{\bf H}_i=\sum_{j_i=1}^{J_i}\beta_{j_i}{\bf a}_{RB}(\theta_{RB,j_i}){\bf a}^H_{Ra}(\theta_{Ra,n_i},\phi_{Ra,n_i}),
\end{align}
where $\beta_{j_i}$ denotes the channel gain of the $j_i$-th path.
\subsection{Received Signal Model}
 {Accordingly, we can present the received signal model in this subsection. }Let ${\bf e}_{t,i} \in \mathbb{C}^{N_{y,z}\times 1}$ denote the phase shift vector of the $i$-th RIS at time slot $t$,
 which satisfies $|[{\bf e}_{t,i} ]_{n_{y,z}}|=1$ for $1\leq n_{y,z} \leq N_{y,z}$.  Accordingly, the received signal from the MU via the $i$-th RIS to the BS at time slot $t$  could be expressed as
\begin{align}\label{v38}
\mathbf{y}(t) = (\sum_{i=1}^{M}{\bf H}_i{\rm Diag}({\bf e}_{t,i} ){\bf h}_i+{\bf h}_{BU})\sqrt{p}s(t)+{\bf n}(t),
\end{align}
where $s(t)$ denotes the transmitted pilot signal of the MU and ${\bf n}(t)\in\mathbb{C}^{N_b\times1}$ is additive white Gaussian
 noise (AWGN) following the distribution of $\mathcal{CN}(0,\delta^2{\bf I})$.
 Moreover, $p$ is the transmit power of the MU.

 \section{Two Step 3D Positioning Scheme}\label{frame}
 {In this paper, the framework of the classical two-step 3D localization scheme is adopted in this paper to estimate the position of the MU.} Specifically, the first step is to estimate the channel parameters (AoAs and TDoAs) from the received signal $\mathbf{y}(t)$ in \eqref{v38}, and the second step is to estimate  the 3D position of the MU according to the estimated channel parameters.
 \subsection{Step I : Channel Parameters Estimation}\label{RPP}
 {Within this subsection, we estimate the channel parameters for localization.} Specifically, we employ the 2D-DFT algorithm \cite{Zhou1} to estimate the AoAs associated with the RISs. Simultaneously, we estimate the TDoA at the RISs through the transmission of a pilot signal via each RIS to the BS. Such a procedure facilitates the computation of the ToAs for both the MU-RIS and MU-RIS-BS links. Thereafter, we describe a geometric relationship correlating the channel parameters with the 3D coordinates of the MU. In the concluding part of this section, a systematic modeling of the estimation, inclusive of its inherent error, is presented.
 \subsubsection{AoA Estimation}
 {According to \cite{wt1}, by employing the 2D-DFT algorithm, the AoAs at the RISs is estimated.} Specifically,  the estimated azimuth and elevation AoAs at the $i$-th RIS of the LoS path \footnote{In the mmWave band, the contributions of the NLoS path components to the channel are minimal, given their rapid variations. Consequently, this paper concentrates on estimating the AoA of the LoS path, which sufficiently informs the design of a subsequent localization algorithm.}   can be given as
    \begin{align}\label{1em18}
&\hat{\theta}_{MR,i}=\arcsin\bigg(\frac{\lambda_c b_{n_i}}{N_z d_r }-\frac{\lambda_c\Tilde{\varpi}_{2,i}}{2\pi d_r}\bigg),\nonumber\\
&\hat{\phi}_{MR,i} =\nonumber\\
& \arccos((\frac{\lambda_c b_{n_i}}{N_y d_r }-\frac{\lambda_c\Tilde{\varpi}_{1,i}}{2\pi d_r})\bigg/\sqrt{(1-(\frac{\lambda_c q_{n_i}}{N_z d_r }-\frac{\lambda_c\Tilde{\varpi}_{2,i}}{2\pi d_r})^2)}),
\end{align}
where $b_{n_i}=\frac{N_y d_r\cos{\theta}_{MR,i}\cos{\phi}_{MR,i}}{\lambda_c}$, $q_{n_i}=\frac{N_z d_r\sin{\phi}_{MR,i}}{\lambda_c}$. Besides, $\Tilde{\varpi}_{1,i}\in[-\pi/N_y,\pi/N_y]$ and $\Tilde{\varpi}_{2,i}\in[-\pi/N_z,\pi/N_z]$  denote the angle rotation parameters.

Consequently, let us introduce the geometry relationship between the AoA information and the position of the MU. For the MU-RIS-BS link, the available angle information includes the AoAs (azimuth and elevation angles) at the RISs, the AODs (azimuth and elevation angles) at the RISs, and the AoAs at the BS. For the MU-BS link, the angle information for localization is the AoA at the BS. However, since we can only estimate the azimuths of the AoAs at the BS due to the assumption of ULAs, the 3D coordinates of the MU cannot be estimated without the elevation of the AoAs. Therefore, we use the AoAs at the RISs for the localization estimation in this paper. Moreover, as NLoS path component usually varies fast and its weight to the channel is marginal, especially in the mmWave band, we are more interested in LoS path. Hence, we intend to estimate the path parameter $( \theta_{MR,i},\phi_{MR,i} )$ of the LoS component from the MU to the RISs, which can be used to derive the position of the MU.
	
It is assumed that the azimuth AoA $\theta_{MR,i}$ at the RIS is the angle between the projection of the wave vector on the x-o-y plane and the y-axis as
	\begin{equation}\label{q4}
		\theta_{MR,i}=\arctan \frac{x_q-x_i}{y_q-y_i},
	\end{equation}
	where $\theta_{MR,i}\in(0,\pi)$.

The elevation AoA $\phi_{MR,i}$ at the RIS of the MU-RIS link is
	\begin{equation}\label{q5}
		\phi_{MR,i}=\arctan \frac{z_q-z_i}{\sin\theta_{MR,i}(x_q-x_i)+\cos\theta_{MR,i}(y_q-y_i)},
	\end{equation}
	where $\phi_{MR,i}\in(-\pi/2,\pi/2)$.
 \subsubsection{TDoA Estimation}
To deduce the TDoA at the RISs, an initial estimation of the ToA at the RIS is imperative. This is achieved by focusing solely on the operation of the $i$-th RIS while deactivating its counterparts. In this configuration, the MU dispatches a pilot signal directed towards the BS, enabling the latter to ascertain the ToA corresponding to the MU-RIS-BS linkage. Taking into consideration the fixed spatial coordinates of both the BS and the RIS, the ToA affiliated with the RIS-BS link can be computed by dividing the spatial separation by the universally constant speed of light. Building upon this foundation, the ToA pertinent to the MU-RIS link is extrapolated by subtracting the previously computed ToA of the RIS-BS link from that of the MU-RIS-BS link. Finally, the desired TDoA at the RISs is derived by subtracting the MU-BS link's ToA from the MU-RIS link's ToA.

Subsequently, we describe the geometric correlation between the TDoA data and the spatial coordinates of the MU. It should be noted that the disparity in link distances can be efficiently derived by scaling the speed of light with the TDoAs. Thus, our subsequent discussions will concentrate on these distance differentials, as they are synonymous with the TDoAs in context.
		
Let $R_{BU}$ denote the true distance of the MU-BS link, which can be calculated as
\begin{equation}\label{t1}
			R_{BU}= \sqrt{(x_q-x_p)^2+(y_q-y_p)^2+(z_q-z_p)^2}.
\end{equation}
Additionally, the true distance from the MU to the $i$-th RIS can be expressed as
		\begin{equation}\label{t2}
			R_{RU,i}= \sqrt{(x_q-x_i)^2+(y_q-y_i)^2+(z_q-z_i)^2}.
		\end{equation}
		Using $R_{BU}$ as the reference distance, the true distance difference between $R_{{RU,i}}$ and $R_{{BU}}$ can be written as
		\begin{equation}\label{t3}
			R_{{B,i}}=R_{{RU,i}}-R_{{BU}} .
		\end{equation}
		The true distance difference $R_{B,i}$ will be used in the following localization algorithm design.
\subsubsection{Estimation of AoAs and TDoAs}
	We have the estimated AoAs and TDoAs modeled as
		\begin{align}\label{new2}
		\hat{\theta}_{MR,i}&={\theta}_{MR,i}+n_i,\nonumber\\
\hat{\phi}_{MR,i}&={\phi}_{MR,i}+\omega_i,\nonumber\\
			\hat{R}_{B,i}&=R_{{B,i}}+\nu_i,
		\end{align}
		where $\{\hat{\theta}_{MR,i},\hat{\phi}_{MR,i},\hat{R}_{B,i}\}$ denote the estimation of $\{{\theta}_{MR,i},{\phi}_{MR,i},R_{{B,i}}\}$ and $\{n_i, \omega_i,\nu_i\}$ denote the estimation error.
		For the sake of illustration, we collect all the estimation localization parameters in the following vectors
		\begin{align}\label{3}
									\setlength{\abovedisplayskip}{1pt}
			\setlength{\belowdisplayskip}{1pt}
			\hat{{\bm \theta}}_{\textrm{r}}={\bm \theta}_{\textrm{r}}+\mathbf{n},\quad
			\hat{\bm{\phi}}_{\textrm{r}}=\bm{\phi}_{\textrm{r}}+{\bm \omega},\quad
			\hat{\bf R}_{B}= {\bf R}_{B}+{\bm \nu},
		\end{align}
		where
\begin{flalign}\label{new3}
    &\hat{\boldsymbol{\theta}}_{\mathrm{r}} = \mathrm{Vec}(\hat{\theta}_{MR}), & &\boldsymbol{\theta}_{\mathrm{r}} = \mathrm{Vec}(\theta_{MR}), & \nonumber \\
    &\hat{\boldsymbol{\phi}}_{\mathrm{r}} = \mathrm{Vec}(\hat{\phi}_{MR}), & &\boldsymbol{\phi}_{\mathrm{r}} = \mathrm{Vec}(\phi_{MR}), & \nonumber \\
    &\hat{\mathbf{R}}_{B} = \mathrm{Vec}(\hat{R}_B), & &\mathbf{R}_{B} = \mathrm{Vec}(R_B), & \nonumber \\
    &\mathbf{n} = \mathrm{Vec}(n), & &\boldsymbol{\omega} = \mathrm{Vec}(\omega), & &\boldsymbol{\nu} = \mathrm{Vec}(\nu), &
\end{flalign}
with $\mathrm{Vec}(x)$ represents the vector $[x_1, x_2, \dots, x_M]^T$.

In prevailing literature, the parameters $\mathbf{n}$, $\bm{\omega}$ and ${\bm\nu}$  are commonly assumed to represent additive zero-mean complex Gaussian noise for the sake of mathematical tractability  \cite{2002A}. While estimating the TDoA at the RISs, both the time delay and the transmitted pilot signal inherently exhibit Gaussian randomness, leading us to maintain the assumption that ${\bm\nu}$ signifies the additive zero-mean complex Gaussian noise. However, as indicated by \cite{wt1}, the probability density functions (PDFs) associated with both elevation and azimuth errors display a non-Gaussian property. Building on this, the variances of ${n}_i$, ${\omega}_i$ can be deduced from the variance derivation presented in \cite{wt1}.  Moreover, this allows us to express the covariance matrices as
\begin{align}
    {\bf Q}_n &= \text{diag}\left(\sigma^2_{n_1}, \dots, \sigma^2_{n_M}\right), \nonumber \\
    {\bf Q}_\omega &= \text{diag}\left(\sigma^2_{\omega_1}, \dots, \sigma^2_{\omega_M}\right), \label{cm1} \\
    {\bf Q}_\nu &= \text{diag}\left(\sigma^2_{\nu_1}, \dots, \sigma^2_{\nu_M}\right), \nonumber
\end{align}
where $\sigma^2_{n_i}$, $\sigma^2_{\omega_i}$ and $\sigma^2_{\nu_i}$ denote the variance of $n_i$, $\omega_{i}$ and $\nu_i$, respectively.
\subsection{Step II: Proposed mWLS Algorithm}\label{3DP}
Drawing upon the estimated AoAs, TDoAs, and their accompanying non-Gaussian and Gaussian estimation errors with distinct variances, we introduce an mWLS localization algorithm in this subsection, yielding a closed-form solution for the MU's position.

Initially, pseudolinear equations are formulated using the AoA and TDoA data from the RISs. Based on these equations, and considering the AoAs' non-Gaussian noises with unique variances, a preliminary estimate of the MU's position is obtained. Subsequent equations are then constructed based on this preliminary estimate. The refined position estimation is computed from these equations.
\subsubsection{Pseudolinear equations based on the AoA estimation}
		First, let us derive the pseudolinear equations based on the geometry relationship of AoAs at the $i$-th RIS. From \eqref{q4}, we have
		\begin{equation}\label{5}
			\cos\theta_{MR,i}(x_q-x_i)-\sin\theta_{MR,i}(y_q-y_i) = 0.
		\end{equation}
		By replacing the true value $\theta_{MR,i}$ with the estimated $\hat{\theta}_{MR,i}$, we have
		\begin{equation}\label{6}
			\eta_{\theta_{MR,i}}= \cos\hat{\theta}_{MR,i}(x_q-x_i)-\sin\hat{\theta}_{MR,i}(y_q-y_i),
		\end{equation}
		where $\eta_{\theta_{MR,i}}$ denotes the residual error of $\theta_{MR,i}$  due to the estimation error.
		For the sake of analysis, let $\breve{\mathbf{g}}_{\theta_{MR,i}}=[-\cos\hat{\theta}_{MR,i}$ $,\sin\hat{\theta}_{MR,i},0]^T$. Using the definitions of ${\bf q}$ and ${\bf s}_i$, \eqref{6} can be rewritten as
		\begin{equation}\label{8}	
			\eta_{\theta_{MR,i}}=\breve{\mathbf{g}}_{\theta_{MR,i}}^T{\bf s}_i- \breve{\mathbf{g}}_{\theta_{MR,i}}^T
			\mathbf{q}.
		\end{equation}
		As the angle estimation algorithms, e.g., 2D-DFT, have good performance \cite{Gao1}, \cite{Zhou1},  thus it is reasonable to assume that  the estimation error is very small. Hence, we have $\sin(n_i)\approx n_i$ and $\cos(n_i)\approx 1$, where  $n_i$ is the estimation error of $\theta_{MR,i}$. Consequently, we have the following approximations
		\begin{align}\label{9}
			\sin\hat{\theta}_{MR,i}=\sin(\theta_{MR,i}+n_i)\approx \sin\theta_{MR,i}+n_i\cos\theta_{MR,i},\nonumber  \\
			\cos\hat{\theta}_{MR,i}=\cos(\theta_{MR,i}+n_i)\approx \cos\theta_{MR,i}-n_i\sin\theta_{MR,i}.
		\end{align}
		By substituting  \eqref{9} into \eqref{6}, we have
		\begin{align}\label{11}
			\eta_{\theta_{MR,i}} & \approx(\cos\theta_{MR,i}-n_i\sin\theta_{MR,i})(x_q-x_i) \nonumber  \\ &\quad-(\sin\theta_{MR,i}+n_i\cos\theta_{MR,i})(y_q-y_i) \nonumber \\
			&=-n_iR_{{RU,i}}\cos\phi_{MR,i}.
		\end{align}
		
		In \eqref{11}, \eqref{5} and $(x_q-x_i)\sin\theta_{MR,i}+(y_q-y_i)\cos\theta_{MR,i} =l_{i1}+l_{i2}= R_{{RU,i}}\cos\phi_{MR,i}$. Next, using \eqref{q5}, we have
		\begin{align}\label{new1}
			0 &= \cos{\phi}_{MR,i}(z_q-z_i)-\sin{\phi}_{MR,i}(\sin\theta_{MR,i}(x_q-x_i)\nonumber  \\ &\quad+\cos\theta_{MR,i}(y_q-y_i)).
		\end{align}
		By replacing $\{\phi_{MR,i},\theta_{MR,i}\}$  with $\{\hat{\phi}_{MR,i},\hat{\theta}_{MR,i}\}$, we have
		\begin{align}\label{12}
			\eta_{\phi_{MR,i}}
			&=\cos\hat{\phi}_{MR,i}(z_q-z_i) -\sin\hat{\phi}_{MR,i}\sin\hat{\theta}_{MR,i}(x_q-x_i)\nonumber  \\ &\quad-\sin\hat{\phi}_{MR,i}
			\cos\hat{\theta}_{MR,i}(y_q-y_i),
		\end{align}
		where $\eta_{\phi_{MR,i}}$ is the residual error of $\phi_{MR,i}$.
		For simplicity, we define $\breve{\mathbf{g}}_{\phi_{MR,i}} = [\sin\hat{\phi}_{MR,i}\sin\hat{\theta}_{MR,i},$ $\sin\hat{\phi}_{MR,i}\cos\hat{\theta}_{MR,i}, -\cos\hat{\phi}_{MR,i}]^T$.
		Then, by utilizing the definitions of ${\bf q}$ and ${\bf s}_i$, \eqref{12} can be expressed as
		\begin{equation}\label{14}
			\eta_{\phi_{MR,i}} = \breve{\mathbf {g}}_{\phi_{MR,i}}^T{\bf s}_i-\breve{\mathbf{g}}_{\phi_{MR,i}}^T{\mathbf  q}.
		\end{equation}
		It is assumed that the estimation error is very small \cite{Daill1}, we have $\sin(\omega_i)\approx \omega_i$ and $\cos(\omega_i)\approx 1$, where $\omega_i$ is the estimation error of $\phi_{MR,i}$. As a result, we have the following approximations as
		\begin{align}\label{15}
			\sin\hat{\phi}_{MR,i}=\sin(\phi_{MR,i}+\omega_i)\approx \sin\phi_{MR,i}+\omega_i\cos\phi_{MR,i}, \nonumber \\
			\cos\hat{\phi}_{MR,i}=\cos(\phi_{MR,i}+\omega_i)\approx \cos\phi_{MR,i}-\omega_i\sin\phi_{MR,i}.
		\end{align}
		Then, by substituting \eqref{15} into  \eqref{12}, and performing some mathematical manipulations, we have
		\begin{align}\label{17}
			\eta_{\phi_{MR,i}}
			&\approx (\cos\phi_{MR,i}-\omega_i\sin\phi_{MR,i})(z_q-z_i)\nonumber \\
			&\quad-(\sin\phi_{MR,i}+\omega_i\cos\phi_{MR,i})\nonumber  \\ &\quad((x_q-x_i)\sin\theta_{MR,i}+(y_q-y_i)\cos\theta_{MR,i})\nonumber \\
			&=-\omega_i R_{{RU,i}}.
		\end{align}
		To derive \eqref{17}, we have used \eqref{new1} and $((z_q-z_i)\sin\phi_{MR,i}+((x_q-x_i)\sin{\theta}_{MR,i}+(y_q-y_i)
		\cos{\theta}_{MR,i})$ $\cos\phi_{MR,i} )=d_{i1}+d_{i2}=R_{{RU,i}}$.
		
		Finally,  using \eqref{8}, \eqref{11}, \eqref{14} and \eqref{17}, we have
		\begin{align}\label{18}
			\breve{\mathbf g}_{\theta_{MR,i}}^T\mathbf{s}_i-\breve{\mathbf{g}}_{\theta_{MR,i}}^T\mathbf{q}&\approx-n_iR_{{RU,i}}
			\cos\phi_{MR,i},\nonumber  \\
			\breve{\mathbf{g}}_{\phi_{MR,i}}^T\mathbf{s}_i-\breve{\mathbf{g}}_{\phi_{MR,i}}^T\mathbf{q}&\approx -\omega_i R_{{RU,i}}.
		\end{align}
		\subsubsection{Pseudolinear equations based on the TDoA estimation}
Then, let us derive the pseudolinear equations based on  the TDoA estimation.
		By taking the square of both sides of \eqref{t1}, we have\vspace{-0.1cm}
		\begin{equation}\label{t4}
			R^{2}_{{BU}}=K_q^{2}+K_p^2-2x_qx_p-2y_qy_p-2z_qz_p,
		\end{equation}
		where $K_q^{2}=x_q^{2}+y_q^{2}+z_q^{2}$ and $K_p^2=x_p^2+y_p^2+z_p^2$.
		Similarly, by taking the square of both sides of \eqref{t2}, we have
		\begin{align}\label{t5}
			R^{2}_{{RU,i}}=K_q^{2}+K_i^2-2x_qx_i-2y_qy_i-2z_qz_i,
		\end{align}
		where $K_i^2=x_i^2+y_i^2+z_i^2$. Then,   using \eqref{t3}, we can obtain $R_{{RU,i}}=R_{{B,i}}+R_{{BU}}$. Thus,  we have
		\begin{align}\label{t6}
			R^{2}_{{RU,i}}=(R_{{B,i}}+R_{{BU}})^2.
		\end{align}
		By substituting \eqref{t5} into \eqref{t6} and expanding the right hand side of \eqref{t6}, \eqref{t6} is rewritten as
		\begin{align}\label{t7}
			&R^{2}_{{B,i}}+2R_{{B,i}} R_{{BU}}+R^{2}_{{BU}}\nonumber  \\ &\quad=K_q^{2}+K_i^2-2x_qx_i-2y_qy_i-2z_qz_i.
		\end{align}
	
		Moreover,  using \eqref{t4} and \eqref{t7}, we have the following expression
		\begin{align}\label{t9}
			&R^{2}_{{B,i}}+2R_{{B,i}} R_{{BU}}\nonumber  \\ &
=K_i^2-K_p^2-2x_{{i,p}}x_q-2y_{{i,p}}y_q-2z_{{i,p}}z_q,
		\end{align}
		where $x_{{i,p}}=(x_i-x_p)$, $y_{{i,p}}=(y_i-y_p)$ and $z_{{i,p}}=(z_i-z_q)$.  In order to derive the pseudolinear equations related to TDoA, \eqref{t9} can be transformed into
		\begin{align}\label{t11}
			&-x_{{i,p}}x_q-y_{{i,p}}y_q-z_{{i,p}}z_q-R_{{B,i}} R_{{BU}}\nonumber  \\ &\quad=-\frac{1}{2}K_i^2+\frac{1}{2}K_p^2+\frac{1}{2}R^{2}_{{B,i}}.
		\end{align}
		Additionally, for the sake of analysis, we define
		\begin{align}\label{t12}
			\mathbf{g}_{{t,i}} &=-[x_{{i,p}},y_{{i,p}},z_{{i,p}},R_{{B,i}}]^T,\nonumber  \\
			\mathbf{u} &=[x_q,y_q,z_q,R_{BU}]^T,\nonumber  \\
			h_i &=-\frac{1}{2}K_i^2+\frac{1}{2}K_p^2+\frac{1}{2}R^{2}_{{B,i}}.
		\end{align}
		Accordingly, we can derive the pseudolinear equation with respect to the TDoA  as			
		\begin{align}\label{td15}
			h_i-\mathbf{g}^{T}_{{t,i}}\mathbf{u}=0.
		\end{align}			
		As $R_{B,i}$ is estimated as $\hat{R}_{B,i}$ by applying the TDoA estimation, by  replacing $R_{B,i}$ with $\hat{R}_{B,i}$, $\mathbf{g}_{{t,i}}$ and ${h}_i$ in \eqref{td15} are re-interpreted as $\hat{\mathbf{g}}_{{t,i}}$ and $\hat{h}_i$, given by		
		\begin{align}\label{tt12}
			\hat{\mathbf{g}}_{{t,i}}& =-[x_{{i,p}},y_{{i,p}},z_{{i,p}},\hat{R}_{{B,i}}]^T,\nonumber  \\
			\hat{h}_i&=-\frac{1}{2}K_i^2+\frac{1}{2}K_p^2+\frac{1}{2}\hat{R}^{2}_{{B,i}}.
		\end{align}
		Thus, according to \cite{2002A}, \eqref{td15} can be derived as		
		\begin{align}\label{t15we}
			\eta_t=\hat{h}_i-\hat{\mathbf{g}}_{{t,i}}^T\mathbf{u} = R_{{RU,i}}\nu_{i}+\frac{1}{2}\nu_{i}^2,
		\end{align}
		where $\eta_t$ and $\nu_{i}$ denote the residual error and the TDoA estimation error of ${R}_{B,i}$ in \eqref{new2}, respectively. As $\nu_{i}\ll R_{{RU,i}}$ is always satisfied in practice, the second order term on the right hand side of \eqref{t15we} can be neglected, and \eqref{t15we} can be approximated as			
		\begin{align}\label{t15}
			\hat{h}_i-\hat{\mathbf{g}}_{{t,i}}^T\mathbf{u} \approx R_{{RU,i}}\nu_{i}.
		\end{align}
\subsubsection{ Preliminary Estimation}
Then, let us derive the preliminary estimation based on the pseudolinear equations \eqref{18} and \eqref{t15}, both of which contain the unknown location ${\bf q}$.

Hence, by combining \eqref{18} and \eqref{t15}, we can derive the following compact form of equations as			\begin{equation}\label{20}
			\hat{\mathbf{h}}_{\textrm{r}}-\hat{\mathbf{G}}_{\textrm{r}}\mathbf{u} = \mathbf{B}_{\textrm{r}}\mathbf{z}_{\textrm{r}},
		\end{equation}
		where $\mathbf{z}_{\textrm{r}}=[\mathbf{n}^T,\bm{\omega}^T,\bm{\nu}^T]^T$ and its covariance matrix is written as
\begin{equation}\label{21}
	\mathbf{Q}_{\textrm{r}} = \text{diag}(\mathbf{Q}_{n}, \mathbf{Q}_{\omega}, \mathbf{Q}_{\nu}).
\end{equation}
		On the left hand side of \eqref{20}, we have		
		\begin{align}\label{22}		
		\hat{\mathbf{h}}_{\textrm{r}}&=[\mathbf{1}^T(\breve{\mathbf{G}}_{\theta_{\textrm{r}}}\odot\mathbf{S})^T,
			\mathbf{1}^T(\breve{\mathbf{G}}_{\phi_{\textrm{r}}}\odot\mathbf{S})^T,\hat{\mathbf{h}}^T_{{t}}]^T,\nonumber\\
			\hat{ \mathbf{G}}_{\textrm{r}}&=[\hat{\mathbf{G}}^T_{\theta_{\textrm{r}}},\hat{\mathbf{G}}^T_{\phi_{\textrm{r}}},\hat{\mathbf{G}}^T_{{t}}]^T,
		\end{align}
		where		
		\begin{align}\label{t24}		\breve{\mathbf{G}}_{\theta_\textrm{r}}&=[\breve{\mathbf{g}}_{\theta_{{MR,1}}},\breve{\mathbf{g}}_{\theta_{{MR,2}}}
			,\cdots,\breve{\mathbf{g}}_{\theta_{{MR,M}}}]^T,\nonumber\\
			\breve{\mathbf{G}}_{\phi_\textrm{r}}&=[\breve{\mathbf{g}}_{\phi_{{MR,1}}},\breve{\mathbf{g}}_{\phi_{{MR,2}}}
			,\cdots,\breve{\mathbf{g}}_{\phi_{{MR,M}}}]^T,\nonumber\\
			\hat{\mathbf{G}}_{\theta_\textrm{r}}&=[\hat{\mathbf{g}}_{\theta_{{MR,1}}},\hat{\mathbf{g}}_{\theta_{{MR,2}}}
			,\cdots,\hat{\mathbf{g}}_{\theta_{{MR,M}}}]^T,\nonumber\\
			\hat{\mathbf{G}}_{\phi_\textrm{r}}&=[\hat{\mathbf{g}}_{\phi_{{MR,1}}},\hat{\mathbf{g}}_{\phi_{{MR,2}}}
			,\cdots,\hat{\mathbf{g}}_{\phi_{{MR,M}}}]^T,\nonumber\\
			\hat{\mathbf{h}}_{{t}}&=[\hat{h}_{{1}},\hat{h}_{{2}}
			,\cdots,\hat{h}_{{M}}]^T,\quad
			\mathbf{S}=[\mathbf{s}_{{1}},\mathbf{s}_{{2}},\cdots,\mathbf{s}_{{M}}]^T,\nonumber\\
			\hat{\mathbf{G}}_{{t}}&=[\hat{\mathbf{g}}_{{t,1}},\hat{\mathbf{g}}_{{t,2}}
			,\cdots,\hat{\mathbf{g}}_{{t,M}}]^T,\nonumber\\
\hat{\mathbf{g}}_{\theta_{MR,i}}&=[\breve{\mathbf{g}}^T_{\theta_{MR,i}},0]^T,\quad
			\hat{\mathbf{g}}_{\phi_{MR,i}}=[\breve{\mathbf{g}}^T_{\phi_{MR,i}},0]^T.
		\end{align}
		On the right hand side of \eqref{20}, we have
		\begin{align}\label{26}
			\mathbf{B}_{\textrm{r}} & =[\mathbf{B}_\textrm{r,n},\mathbf{B}_{\textrm{r,}\omega},\mathbf{B}_{\textrm{r,}\nu}]^T,
		\end{align}
		where			
		\begin{align}\label{new26}
			& \mathbf{B}_\textrm{r,n} =[\mathbf{B}_{\theta_{{MR,n}}},\mathbf{O},\mathbf{O}]^T ,\nonumber\\
	&\mathbf{B}_{{\textrm{r},}\omega}=[\mathbf{O},\mathbf{B}_{\phi_{{MR,}\omega}},\mathbf{O}]^T,\nonumber\\
			&\mathbf{B}_{{\textrm{r},}\nu}=[\mathbf{O},\mathbf{O},\mathbf{B}_{{t}} ]^T,\nonumber\\
			&  \mathbf{B}_{\theta_{{MR,n}}}  =-\textrm{diag}[R_{{RU,1}}\cos\phi_{{MR,1}},\cdots,R_{{RU,1}}\cos\phi_{{MR,M}}],\nonumber\\
			&   \mathbf{B}_{\phi_{{MR,}\omega}} =-\textrm{diag}[R_{RU,1},\cdots,R_{RU,M}],\nonumber\\
			&\mathbf{B}_{t} = \textrm{diag}[R_{RU,1},\cdots,R_{RU,M}].
		\end{align}

		Based on \eqref{20}, the WLS cost function can be formulated as		
		\begin{align}\label{f1}
			f( {\mathbf{u}}) &=  (\hat{{\bf h}}_{\textrm{r}}-\hat{\mathbf{G}}_{\textrm{r}}{\mathbf{u}})^T\mathbf{W}_{\textrm{r}}(\hat{\mathbf{h}}_{\textrm{r}}
			-\hat{\mathbf{G}}_{\textrm{r}}{\mathbf{u}})\nonumber\\
			&= \hat{\mathbf{h}}_{\textrm{r}}^T\mathbf{W}_{\textrm{r}}\hat{\mathbf{h}}_{\textrm{r}}
			-2\hat{\mathbf{h}}_{\textrm{r}}^T\mathbf{W}_{\textrm{r}}\hat{\mathbf{G}}_{\textrm{r}}{\mathbf{u}}
			+{\mathbf{u}}^T\hat{\mathbf{G}}_{\textrm{r}}^T\mathbf{W}_{\textrm{r}}\hat{\mathbf{G}}_{\textrm{r}}{\mathbf{u}},
		\end{align}
		where   $\mathbf{W}_\textrm{r}$ denotes the weight matrix of preliminary estimation.
		
		To derive the estimation of ${\bf u}$, the cost function $f( {\mathbf{u}})$ in \eqref{f1} should be minimized. By setting the first-order derivative of $f( {\mathbf{u}})$ equal to zero, we have 	$\frac{\partial f( {\mathbf{u}})}{\partial{\mathbf{u}}}
		=-2(\hat{\mathbf{G}}_{\textrm{r}}^T\mathbf{W}_{\textrm{r}}\hat{\mathbf{h}}_{\textrm{r}})
		+2(\hat{\mathbf{G}}_{\textrm{r}}^T\mathbf{W}_{\textrm{r}}\hat{\mathbf{G}}_{\textrm{r}}{\mathbf{u}})= 0$. Hence, $\mathbf{u}$ is estimated as		
		\begin{equation}\label{28}
			\breve{\mathbf{u}} = [\breve{x}_q,\breve{y}_q,\breve{z}_q,\breve{R}_{BU}]^T=(\hat{\mathbf{G}}_{\textrm{r}}^T\mathbf{W}_{\textrm{r}}\hat{\mathbf{G}}_{\textrm{r}})^{-1}
			\hat{\mathbf{G}}_{\textrm{r}}^T\mathbf{W}_{\textrm{r}}\hat{\mathbf{h}}_{\textrm{r}}.
		\end{equation}
		As the estimation error is correlated, the weight matrix $\mathbf{W}_{\textrm{r}}$ should be equal to the inverse of  the covariance matrix of the estimation error as		
		\begin{equation}\label{re29}
			\mathbf{W}_{\textrm{r}} = {\bm \Omega}_{\textrm{r}}^{-1},
			{\bm \Omega}_{\textrm{r}} =\mathbb{E}[\mathbf{B}_{\textrm{r}}\mathbf{z}_{\textrm{r}}\mathbf{z}_{\textrm{r}}^T\mathbf{B}_{\textrm{r}}^{T}]
			=\mathbf{B}_{\textrm{r}}\mathbf{Q}_{\textrm{r}}\mathbf{B}_{\textrm{r}}^{T},
		\end{equation}
		where ${\bm \Omega}_{\textrm{r}}$ denotes the covariance matrix of the estimation error.
		However, as shown in \eqref{26} and \eqref{new26}, matrix $\mathbf{B}_{\textrm{r}}$ contains the true distances $\{R_{{RU,1}},R_{{RU,2}},\cdots,R_{{RU,M}}\}$, which remain unknown. Therefore, the weight matrix is initialized by taking the estimated distances as the approximation of  true values. Then, the weight matrix can be iteratively updated by the new estimations.
		\subsubsection{Refining Estimation}
		As we can see from \eqref{t1}, $R_{BU}$ and  ${\bf q}=[x_q,y_q,z_q]^T$ are related, we aim to utilize this relationship to  construct a new set of pseudolinear equations to derive the refining estimation. Considering the estimation error, the relationship between the MU's preliminary estimated position and the BS's position is 			
		\begin{equation}\label{s2_2}
			\left[ {\begin{array}{*{20}{c}}
					(\breve{x}_q-x_p)^2,
					(\breve{y}_q-y_p)^2,
					(\breve{z}_q-z_p)^2,
					\breve{R}_{BU}^2
			\end{array}} \right]^T  \buildrel \Delta \over =\hat{{\bf h}}_{1}.
		\end{equation}
		Without the estimation error, the relationship between the MU's position and the BS's position is represented by 			
		\begin{equation}\label{sbb2_2}
			\left[ {\begin{array}{*{20}{c}}
					({x}_q-x_p)^2,
					({y}_q-y_p)^2,
					({z}_q-z_p)^2,
					{R}_{BU}^2
			\end{array}} \right]^T  \buildrel \Delta \over ={\bf G}_1 {\bm \xi},
		\end{equation}
		where 			
		\begin{align}\label{s2_3}
			{\bf G}_1  &= \left[ {\begin{array}{*{20}{c}}
					{\bf I}_{3\times 3}\\
					{\bf 1}_{1\times 3}
			\end{array}} \right],\nonumber\\
			{\bm \xi}  &=\left[ {\begin{array}{*{20}{c}}
					(x_q-x_p)^2,
					(y_q-y_p)^2,
					(z_q-z_p)^2
			\end{array}} \right]^T\nonumber\\
&=\left({\bf q}-{\bf p}\right)\odot\left({\bf q}-{\bf p}\right).
		\end{align}
		Using \eqref{s2_2} and \eqref{sbb2_2},  we can derive the compact form of the pseudolinear equations as		\begin{equation}\label{s2_5}
			\hat{{\bf h}}_1-{\bf G}_1 {\bm \xi}={\bf z}_1,
		\end{equation}
		where ${\bf z}_1=[z_{11},z_{12},z_{13},z_{14}]^T=[(\breve{x}_q-x_p)^2,(\breve{y}_q-y_p)^2,(\breve{z}_q-z_p)^2,
		\breve{R}_{BU}^2]^T-[({x}_q-x_p)^2,({y}_q-y_p)^2,({z}_q-z_p)^2,{R}_{BU}^2]^T$ denotes the residual vector due to the estimation error. Denote the localization estimation error of the preliminary estimation as $\tilde{\bf u}= [e_1,e_2,e_3,e_4]^T$, i.e ., $\breve{{\bf u}} = {{\bf u}} + \tilde{\bf u}$.
		Then, the elements of ${\bf z}_1$ can be represented as the functions of the elements of error $\tilde{\bf u}$,  which can be expressed as
		\begin{align}\label{s2_6}
		z_{11} &=(\breve{x}_q-x_p)^2-(x_q-x_p)^2\approx 2(x_q-x_p)e_1,\nonumber\\
		z_{12} &=(\breve{y}_q-y_p)^2-(y_q-y_p)^2 \approx 2(y_q-y_p)e_2,\nonumber\\
		z_{13} &=(\breve{z}_q-z_p)^2-(z_q-z_p)^2 \approx 2(z_q-z_p)e_3,\nonumber\\
		z_{14} &= \breve{R}_{BU}^2-R_{BU}^2 \approx 2R_{BU}e_4.
		\end{align}
		Based on \eqref{s2_5}, we can again apply the WLS method to estimate ${\bm \xi}$, thus the WLS cost function can be written as			
		\begin{align}\label{ss2_6}
						\setlength{\abovedisplayskip}{1pt}
			\setlength{\belowdisplayskip}{1pt}
			f({\bm \xi}) &=  (\hat{\bf h}_{1}-\mathbf{G}_{1}{\bm \xi})^T {\mathbf{W}}_{1}(\hat{\mathbf{h}}_{1}
			-\mathbf{G}_{1}{\bm \xi})\nonumber\\&= \hat{\mathbf{h}}_{1}^T{\mathbf{W}}_{1}\hat{\mathbf{h}}_{1}
			-2\hat{\mathbf{h}}_{1}^T{\mathbf{W}}_{1}\mathbf{G}_{1}{\bm \xi}
			+{\bm \xi}^T\mathbf{G}_{1}^T{\mathbf{W}}_{1}\mathbf{G}_{1}{\bm \xi},
		\end{align}
		where  ${\mathbf{W}}_{1}$ denotes the weight matrix of the refining estimation.
		To derive the estimation of ${\bm \xi}$,  $f({\bm \xi})$ in \eqref{ss2_6} should be minimized, leading to $\frac{\partial f( {\bm \xi})}{\partial{\bm \xi}}
		=-2(\mathbf{G}_{1}^T{\mathbf{W}}_{1}{\hat{\mathbf h}}_{1})
		+2(\mathbf{G}_{1}^T{\mathbf{W}}_{1}\mathbf{G}_{1}{\bm \xi})= 0$.
		Therefore,  the estimation of ${\bm \xi}$ is given by		
		\begin{equation}\label{s2_7}
						\setlength{\abovedisplayskip}{3pt}
			\setlength{\belowdisplayskip}{3pt}
			\breve{{\bm \xi}} =  ({\bf G}_1^T{\mathbf{W}}_1{\bf G}_1)^{-1}{\bf G}^T_1{\bf W}_1\hat{\bf h}_1.
		\end{equation}			
		By defining ${\bm \Omega}_1 =\mathbb{E}[{\bf z}_1{\bf z}_1^T]$ as the covariance matrix of ${\bf z}_1$,  we have  $\mathbf{W}_1 = {\bm \Omega}^{-1}_1 $. Moreover, ${\bm \Omega}_1$ can be further derived as
		\begin{equation}\label{s2_8}
						\setlength{\abovedisplayskip}{1pt}
			\setlength{\belowdisplayskip}{1pt}
			{\bm \Omega}_1 ={{\bf B}}_1 (\hat{{\bf G}}_\textrm{r}^T{\bf W}_\textrm{r}\hat{{\bf G}}_\textrm{r})^{-1}{\mathbf{B}}_1 ,
			\quad{{\bf B}}_1 = 2 \textrm{diag} \left\{{{\bf u}}-\left[ {\begin{array}{*{20}{c}}
					{\bf p}\\
					0
			\end{array}} \right]\right\},
		\end{equation}
		where the proof of \eqref{s2_8} is given in Appendix A of \cite{TWu}. Although ${\bf B}_1$ contains the true values $\{x_q,y_q,z_q,R_{BU}\}$, it can be approximated as
		\begin{equation}\label{ab7}
						\setlength{\abovedisplayskip}{1pt}
			\setlength{\belowdisplayskip}{1pt}
			\hat{\bf B}_1=2 \textrm{diag} \left\{\breve{{\bf u}}-\left[ {\begin{array}{*{20}{c}}
					{\bf p}\\
					0
			\end{array}} \right]\right\}.
		\end{equation}
		Then, ${\bm \Omega}_1$ can be approximated as $\hat{\bm \Omega}_1= \hat{\bf B}_1 (\hat{{\bf G}}_{\textrm{r}}^T{{\bf W}}_{\textrm{r}}\hat{{\bf G}}_{\textrm{r}})^{-1}\hat{\bf B}_1$. Hence, ${\bf W}_1$ can be estimated as $\hat{\bf W}_1 =\hat{\bm \Omega}_1^{-1}= \hat{\bf B}_1 ^{-1}(\hat{{\bf G}}_{\textrm{r}}^T{{\bf W}}_{\textrm{r}}\hat{{\bf G}}_{\textrm{r}})\hat{\bf B}_1^{-1}$, and $\breve{{\bm \xi}}$ is approximated as
		\begin{equation}\label{s2_7}
			\breve{{\bm \xi}} \approx  ({\bf G}_1^T\hat{\mathbf{W}}_1{\bf G}_1)^{-1}{\bf G}^T_1\hat{\bf W}_1\hat{\bf h}_1.
		\end{equation}
Finally, by using the estimation $\breve{\mathbf{u}}$ in \eqref{28} and $\breve{{\bm \xi}}$ in \eqref{s2_7}, the final closed-form estimation of the position of MU can be expressed as:
		\begin{equation}\label{s3_1}
			\setlength{\abovedisplayskip}{3pt}
			\setlength{\belowdisplayskip}{3pt}
			\hat{{\bf q}} = {\bm \Pi}\sqrt{\breve{{\bm \xi}}}+{\bf p},\quad
			{\bm \Pi} = \textrm{diag}\{\textrm{sgn}(\breve{\mathbf{u}}(1:3)-{\bf p})\},
		\end{equation}
		where $\textrm{sgn}(\textrm{x})$ denotes the signum function.

\section{Bias Analysis} \label{biasa}
In prevailing literatures \cite{Henk}, the CRLB is typically applied to Gaussian-distributed data for localization performance. However, in this paper, the estimation errors of the  AoAs are non-Gaussian, posing challenges with CRLB analysis. To circumvent this, we employ bias analysis \cite{2012Bias}, a simpler measure of an estimator's systematic error, offering a direct insight into primary inaccuracies compared to RMSE.

\subsection{Decomposing Estimation Error Expression}
Given that the bias is determined by computing the expectation of the estimation error, and considering the error terms are interlinked within the expression, it becomes essential to decompose the expression to accurately derive the expectation of the error.
				
As the estimation error is inevitable,  $\hat{\mathbf{G}}_{\textrm{r}}$ defined in \eqref{22} consists of two parts: the true matrix ${\bf G}_{\textrm{r}}$ and the error matrix $\tilde{\bf{G}}_{\textrm{r}}$, i.e., $\hat{\mathbf{G}}_{\textrm{r}}=\mathbf{G}_{\textrm{r}}+\tilde{\mathbf{G}}_{\textrm{r}}$.
				According to \eqref{22} and \eqref{t24},  to derive the expressions of $\mathbf{G}_{\textrm{r}}$ and $\tilde{\mathbf{G}}_{\textrm{r}}$, we need to decompose $\hat{\mathbf{g}}_{\theta_{MR,i}}$,  $\hat{\mathbf{g}}_{\phi_{MR,i}}$ and $\hat{\mathbf{g}}_{t,i}$ into the true values and the estimation error at first.

				First, let us decompose $\hat{\mathbf{g}}_{\theta_{MR,i}}$ into the true value and the estimation error.
				Using the approximations in \eqref{9} and the definition of $\hat{\mathbf{g}}_{\theta_{MR,i}}$ in \eqref{t24}, $\hat{\mathbf{g}}_{\theta_{MR,i}}$ is approximated as $\hat{\mathbf{g}}_{\theta_{MR,i}}\approx [-\cos\theta_{MR,i},\sin\theta_{MR,i},0,0]^T
				+n_i[\sin\theta_{MR,i},\cos\theta_{MR,i},0,0]^T$.
				By defining $\bar{\mathbf{g}}_{\theta_{MR,i}}=[-\cos\theta_{MR,i},$  $\sin\theta_{MR,i},0,0]^T$ and ${\bf g}_{1n_{i}}=[\sin\theta_{MR,i},\cos\theta_{MR,i},0,0]^T$, we have
				\begin{equation}\label{b2}
					\setlength{\abovedisplayskip}{3pt}
					\setlength{\belowdisplayskip}{3pt}
					\hat{\mathbf{g}}_{\theta_{MR,i}}\approx\bar{\mathbf{g}}_{\theta_{MR,i}}+n_i{\bf g}_{1n_{i}},
				\end{equation}
				where $n_i{\bf g}_{1n_{i}}$ denotes the error term.

				Similarly, by utilizing the approximations in \eqref{9}, \eqref{15} and the definition of $\hat{\mathbf{g}}_{\phi_{MR,i}}$ in \eqref{t24}, $\hat{\mathbf{g}}_{\phi_{MR,i}}$ is approximated  as
				\begin{align}
					\begin{aligned}
						\setlength{\abovedisplayskip}{1pt}
						\setlength{\belowdisplayskip}{1pt}
					&\hat{\mathbf{g}}_{\phi_{MR,i}}\nonumber \\
&\approx[(\sin\phi_{MR,i}+\omega_i\cos\phi_{MR,i})(\sin\theta_{MR,i}+n_i\cos\theta_{MR,i}),\nonumber \\
					&\quad(\sin\phi_{MR,i}+\omega_i\cos\phi_{MR,i})(\cos\theta_{MR,i}-n_i\sin\theta_{MR,i}),\nonumber\\
					&\quad-(\cos\phi_{MR,i}-\omega_i\sin\phi_{MR,i}),0]^T.\nonumber
					\end{aligned}
				\end{align}
				By ignoring the terms $\omega_in_i\cos\phi_{MR,i}\cos\theta_{MR,i}$ and $-\omega_in_i\cos\phi_{MR,i}\sin\theta_{MR,i}$ \cite{AoA} and defining $\bar{\mathbf{g}}_{\phi_{MR,i}}=\left[\sin\phi_{MR,i}\sin\theta_{MR,i},
				\sin\phi_{MR,i}\cos\theta_{MR,i},-\cos\phi_{MR,i},0\right]^T$, $\mathbf{g}_{\omega_{i}}=[\cos\phi_{MR,i}$ $\sin\theta_{MR,i}
				,$ $\cos\phi_{MR,i}\cos\theta_{MR,i},\sin\phi_{MR,i},0]^T$ and $\mathbf{g}_{2n_{i}}=[\sin\phi_{MR,i}\cos\theta_{MR,i}
				,$ $-\sin\phi_{MR,i}\sin\theta_{MR,i},$ $0,0]^T$, we have
				\begin{equation}\label{b5}
					\setlength{\abovedisplayskip}{2pt}
					\setlength{\belowdisplayskip}{2pt}
					\hat{\mathbf{g}}_{\phi_{MR,i}} \approx\bar{\mathbf{g}}_{\phi_{MR,i}}+\omega_i\mathbf{g}_{\omega_{i}}+n_i\mathbf{g}_{2n_{i}},
				\end{equation}
				where $\omega_i\mathbf{g}_{\omega_{i}}$ and $n_i\mathbf{g}_{2n_{i}}$ denote the error terms, respectively.
				
				Similarly,  we can decompose $\hat{\mathbf{g}}_{t,i}$ into the true value and the estimation error. Based on the definition of $\mathbf{g}_{t,i}$ in \eqref{t12} and $\hat{\mathbf{g}}_{t,i}$ in \eqref{tt12},  we can rewrite  $\hat{\mathbf{g}}_{t,i}$ as
					$\hat{\mathbf{g}}_{t,i}
					=-[x_{{i,p}},y_{{i,p}},z_{{i,p}},R_{{B,i}}]^T-[0,0,0,\nu_i]^T
					=\mathbf{g}_{t,i}+\nu_i[0,0,0,-1]^T$.
				Letting $\mathbf{g}_{\nu} = [0,0,0,-1]^T$, we have
												\vspace{-0.1cm}
				\begin{align}\label{bb6}
					\hat{\mathbf{g}}_{t,i}=\mathbf{g}_{t,i}+\nu_i\mathbf{g}_{\nu},				
				\end{align}				
				where $\nu_i\mathbf{g}_{\nu}$ denotes the error term.
				
				Therefore,  using \eqref{b2}, \eqref{b5} and \eqref{bb6}, $\hat{\mathbf{G}}_{\textrm{r}}$ can be written as
				\begin{equation}
					\setlength{\abovedisplayskip}{3pt}
					\setlength{\belowdisplayskip}{3pt}
					\label{b11}
					\hat{\mathbf{G}}_{\textrm{r}} =\mathbf{G}_{\textrm{r}}+\tilde{\mathbf{G}}_{\textrm{r}}.
				\end{equation}
				Moreover, using the definition of ${\bf n }$, $\bm{\omega}$ and $\bm{\nu}$ in \eqref{new3}, $\tilde{\mathbf{G}}_{\textrm{r}}$ can be written as
				\begin{align}\label{bb18}
					\tilde{\mathbf{G}}_{\textrm{r}}
					=& \bar{\mathbf{G}}_{n}\odot [(\mathbf{n1}_{1\times4})^T,(\mathbf{n1}_{1\times4})^T,(\mathbf{n1}_{1\times4})^T]^T\nonumber \\
					&+\bar{\mathbf{G}}_{\omega}\odot [(\bm{\omega1}_{1\times4})^T,(\bm{\omega1}_{1\times4})^T,(\bm{\omega1}_{1\times4})^T]^T\nonumber \\
					&+\bar{\mathbf{G}}_{\nu}\odot [(\bm{\nu1}_{1\times4})^T,(\bm{\nu1}_{1\times4})^T,(\bm{\nu1}_{1\times4})^T]^T,
				\end{align}
				where				
				\begin{align}\label{bbn19}
					\bar{\mathbf{G}}_{n}&=[\mathbf{G}_{1n}^T,\mathbf{G}_{2n}^T,\mathbf{O}_{4\times M}]^T,\nonumber \\
					\bar{\mathbf{G}}_{\omega}&=[\mathbf{O}_{4\times M},\mathbf{G}_{\omega}^T,\mathbf{O}_{4\times M}]^T,\nonumber \\
					\bar{\mathbf{G}}_{\nu}&=[\mathbf{O}_{4\times M},\mathbf{O}_{4\times M},\mathbf{G}_{\nu}^T]^T,\nonumber \\
					\mathbf{G}_{1n}&=[\mathbf{g}_{1n_{1}},\mathbf{g}_{1n_{2}}
					,\cdots,\mathbf{g}_{1n_{M}}]^T,\nonumber \\
					\mathbf{G}_{2n}&=[\mathbf{g}_{2n_{1}},\mathbf{g}_{2n_{2}}
					,\cdots,\mathbf{g}_{2n_{M}}]^T,\nonumber \\
					\mathbf{G}_{\omega}&=[\mathbf{g}_{\omega_{1}},\mathbf{g}_{\omega_{2}}
					,\cdots,\mathbf{g}_{\omega_{M}}]^T,\nonumber \\
					\mathbf{G}_{\nu}&=[\mathbf{g}_{\nu},\mathbf{g}_{\nu}
					,\cdots,\mathbf{g}_{\nu}]^T.
				\end{align}
				In addition, $\mathbf{G}_{\textrm{r}}$ is given by
				\begin{equation}\label{bb19}
					\setlength{\abovedisplayskip}{3pt}
					\setlength{\belowdisplayskip}{3pt}
					\mathbf{G}_{\textrm{r}} = [\mathbf{G}_{\theta_{\textrm{r}}}^T,\mathbf{G}_{\phi_{\textrm{r}}}^T,\mathbf{G}_{{t}}^T]^T,
				\end{equation}
				where
				\begin{align}\label{bb8}
					\mathbf{G}_{\theta_\textrm{r}}&=[\bar{\mathbf{g}}_{\theta_{MR,1}},\bar{\mathbf{g}}_{\theta_{MR,2}}
					,\cdots,\bar{\mathbf{g}}_{\theta_{MR,M}}]^T,\nonumber \\
					\mathbf{G}_{\phi_\textrm{r}}&=[\bar{\mathbf{g}}_{\phi_{{MR,1}}},\bar{\mathbf{g}}_{\phi_{{MR,2}}}
					,\cdots,\bar{\mathbf{g}}_{\phi_{{MR,M}}}]^T,\nonumber\\
					\mathbf{G}_{t}&=[\mathbf{g}_{{t,1}},\mathbf{g}_{{t,2}}
					,\cdots,\mathbf{g}_{{t,M}}]^T.
				\end{align}

			\vspace{-0.5cm}
\subsection{Bias of Preliminary Estimation}
				In this subsection, we aim to derive the bias of preliminary estimation $\breve{\mathbf{u}}$. To derive the bias of $\breve{\bf u}$, the expression of estimation error of $\breve{\mathbf{u}}$ should be derived, which is given by $\tilde{\bf u}=\breve{\bf u}-{\bf u}$. Then, the bias of $\breve{\bf u}$ is given by taking the expectation of $\tilde{\bf u}$, which is written as $\mathbb{E}(\tilde{\bf u})$.
				Using \eqref{28} and the definition of $\mathbf{u}$, $\tilde{\bf u}$ can be further derived as
	\begin{equation}\label{bi1}
					\Tilde{\mathbf{u}}=\breve{\bf u}-{\bf u}
					=(\hat{\mathbf{G}}_{\textrm{r}}^T\mathbf{W}_{\textrm{r}}\hat{\mathbf{G}}_{\textrm{r}})^{-1}
					\hat{\mathbf{G}}_{\textrm{r}}^T\mathbf{W}_{\textrm{r}}(\hat{\mathbf{h}}_{\textrm{r}}
					-\hat{\mathbf{G}}_{\textrm{r}}\mathbf{u}).
				\end{equation}
				Since we have $\hat{\mathbf{h}}_{\textrm{r}}-\hat{\mathbf{G}}_{\textrm{r}}\mathbf{u}=\mathbf{B}_{\textrm{r}}\mathbf{z}_{\textrm{r}}$ in \eqref{20}, $\tilde{\bf u}$ can be derived as $\tilde{\bf u}=(\hat{\mathbf{G}}_{\textrm{r}}^T\mathbf{W}_{\textrm{r}}\hat{\mathbf{G}}_{\textrm{r}})^{-1}
				\hat{\mathbf{G}}_{\textrm{r}}^T\mathbf{W}_{\textrm{r}}\mathbf{B}_{\textrm{r}}\mathbf{z}_{\textrm{r}}$.
				As we have mentioned above, it is necessary to consider the second order term when analyzing the bias \cite{2012Bias}. However,  we have used \eqref{t15} to derive \eqref{20} rather than \eqref{t15we}, which ignores the  second order term $\frac{1}{2}\nu_{i}^2$. Therefore, to obtain the expression of $\tilde{\bf u}$ and derive the bias of the preliminary estimation, $\mathbf{B}_{\textrm{r}}\mathbf{z}_{\textrm{r}}$ should be extended to $\mathbf{B}_{\textrm{r}}\mathbf{z}_{\textrm{r}}
				+[\mathbf{0}^T,\frac{1}{\sqrt{2}}{\bm \nu}^T]^T\odot[\mathbf{0}^T,\frac{1}{\sqrt{2}}\mathbf{\bm \nu}^T]^T$, where $\mathbf{0}$ is a $2 M\times 1 $ column vector. Then, $\tilde{\bf u}$ can be further derived as $\Tilde{\mathbf{u}} =(\hat{\mathbf{G}}_{\textrm{r}}^T\mathbf{W}_{\textrm{r}}\hat{\mathbf{G}}_{\textrm{r}})^{-1}
				\hat{\mathbf{G}}_{\textrm{r}}^T\mathbf{W}_{\textrm{r}}\bigg(\mathbf{B}_{\textrm{r}}\mathbf{z}_{\textrm{r}}
				+[\mathbf{0}^T,\frac{1}{\sqrt{2}}{\bm \nu}^T]^T\odot[\mathbf{0}^T,\frac{1}{\sqrt{2}}\mathbf{\bm \nu}^T]^T\bigg)$.
				Letting $\mathbf{\bm \eta} = [\mathbf{0}^T,\frac{1}{\sqrt{2}}\bm{\nu}^T]^T$ and $\hat{\mathbf{P}}_\textrm{r} =\hat{\mathbf{G}}_\textrm{r}^T{\mathbf{W}}_\textrm{r}\hat{\mathbf{G}}_\textrm{r}$, $\Tilde{\mathbf{u}}$ can be rewritten as
				\begin{equation}
					\setlength{\abovedisplayskip}{4pt}
					\setlength{\belowdisplayskip}{4pt}
					\label{b14}
					\Tilde{\mathbf{u}} = \hat{\mathbf{P}}^{-1}_{\textrm{r}} \hat{\mathbf{G}}_{\textrm{r}}^T
					\mathbf{W}_{\textrm{r}}(\mathbf{B}_{\textrm{r}}\mathbf{z}_{\textrm{r}}+\mathbf{\bm \eta}\odot\mathbf{\bm\eta}).
				\end{equation}
				According to the definition of $\hat{\mathbf{G}}_\textrm{r}$ in \eqref{b11}, $\hat{\mathbf{P}}_{\textrm{r}}$ can be further derived as
			\begin{align}\label{b15}
					\hat{\mathbf{P}}_{\textrm{r}}
					&=(\mathbf{G}_{\textrm{r}}+\tilde{\mathbf{G}}_{\textrm{r}})^T\mathbf{W}_{\textrm{r}}(\mathbf{G}_{\textrm{r}}
					+\tilde{\mathbf{G}}_{\textrm{r}})\nonumber \\
&=\mathbf{G}^{T}_{\textrm{r}}\mathbf{W}_{\textrm{r}}\mathbf{G}_{\textrm{r}}
					+\tilde{\mathbf{G}}^T_{\textrm{r}}\mathbf{W}_{\textrm{r}}\mathbf{G}_{\textrm{r}}
					+\mathbf{G}^{T}_{\textrm{r}}\mathbf{W}_{\textrm{r}}\tilde{\mathbf{G}}_{\textrm{r}}
					+\tilde{\mathbf{G}}^T_{\textrm{r}}\mathbf{W}_{\textrm{r}}\tilde{\mathbf{G}}_{\textrm{r}}.
				\end{align}
				Here, $\tilde{\mathbf{G}}^T_{\textrm{r}}\mathbf{W}_{\textrm{r}}\tilde{\mathbf{G}}_{\textrm{r}}$  can be neglected\footnote{If we consider $\tilde{\mathbf{G}}^T_{\textrm{r}}\mathbf{W}_{\textrm{r}}\tilde{\mathbf{G}}_{\textrm{r}}$ when deriving the expression of $\tilde{\bf u}$, it will be multiplied by $\mathbf{B}_{\textrm{r}}\mathbf{z}_{\textrm{r}}$, and the terms including $\tilde{\mathbf{G}}^T_{\textrm{r}}\mathbf{W}_{\textrm{r}}\tilde{\mathbf{G}}_{\textrm{r}}$ in the final expression of $\tilde{\bf u}$ is higher than the second order, thus we neglect this term here.},  thus  $\hat{\mathbf{P}}_{\textrm{r}}$ can be approximated as
				\begin{equation}\label{b16}
					\hat{\mathbf{P}}_{\textrm{r}} \approx\mathbf{G}^{T}_{\textrm{r}}\mathbf{W}_{\textrm{r}}\mathbf{G}_{\textrm{r}}
					+\tilde{\mathbf{G}}^T_{\textrm{r}}\mathbf{W}_{\textrm{r}}\mathbf{G}_{\textrm{r}}
					+\mathbf{G}^{T}_{\textrm{r}}\mathbf{W}_{\textrm{r}}\tilde{\mathbf{G}}_{\textrm{r}} =\mathbf{P}_{\textrm{r}}+\tilde{\mathbf{P}}_{\textrm{r}},
				\end{equation}
				where $\mathbf{P}_\textrm{r} =\mathbf{G}_\textrm{r}^{T}\mathbf{W}_\textrm{r}\mathbf{G}_\textrm{r}$  and $\tilde{\mathbf{P}}_{\textrm{r}}=\tilde{\mathbf{G}}^T_{\textrm{r}}\mathbf{W}_{\textrm{r}}\mathbf{G}_{\textrm{r}}
				+\mathbf{G}^{T}_{\textrm{r}}\mathbf{W}_{\textrm{r}}\tilde{\mathbf{G}}_{\textrm{r}}$,  respectively. However, if we use the definition of $\hat{\mathbf{P}}_\textrm{r}$ in \eqref{b16} to derive \eqref{b14},  the inverse matrix $\hat{\mathbf{P}}_\textrm{r}^{-1}=(\mathbf{P}_{\textrm{r}}+\tilde{\mathbf{P}}_{\textrm{r}})^{-1}$  is complex and challenging to derive. Fortunately, we can derive the approximation of $\hat{\mathbf{P}}_\textrm{r}^{-1}$ by utilizing the Newmann expansion \cite{2012Bias} when the error level is small, which is written as
				\begin{equation}\label{b18}
					\hat{\mathbf{P}}_\textrm{r}^{-1} \approx\left(\mathbf{I}-\mathbf{P}_\textrm{r}^{-1}\tilde{\mathbf{P}}_\textrm{r}\right)\mathbf{P}_\textrm{r}^{-1}.
				\end{equation}
				By substituting  \eqref{b18} and  \eqref{b11} into \eqref{b14}, $\Tilde{\mathbf{u}}$ is derived as
				\begin{equation}\label{b19}
					\Tilde{\mathbf{u}} \approx \left(\mathbf{I}-\mathbf{P}_{\textrm{r}}^{-1}\tilde{\mathbf{P}}_{\textrm{r}}\right)\mathbf{P}_{\textrm{r}}^{-1}
					(\mathbf{G}_{\textrm{r}}+\tilde{\mathbf{G}}_{\textrm{r}})^T\mathbf{W}_{\textrm{r}}
					(\mathbf{B}_{\textrm{r}}\mathbf{z}_{\textrm{r}}+\bm{\eta}\odot\bm{\eta}).
				\end{equation}
				For the sake of illustration, using $\mathbf{P}_{\textrm{r}}$ in \eqref{b16}, let us define $\mathbf{H}_{\textrm{r}} =(\mathbf{G}_{\textrm{r}}^{T}\mathbf{W}_{\textrm{r}}\mathbf{G}_{\textrm{r}})^{-1}
				\mathbf{G}_{\textrm{r}}^{T}\mathbf{W}_{\textrm{r}}= \mathbf{P}^{-1}_{\textrm{r}}\mathbf{G}_{\textrm{r}}^{T}\mathbf{W}_{\textrm{r}}$.
				Then, $\Tilde{\mathbf{u}}$ can be represented as
				\begin{align}\label{sssb19}
						\hspace{-0.5cm}\Tilde{\mathbf{u}}
					&\approx\mathbf{H}_{\textrm{r}}\mathbf{B}_{\textrm{r}}\mathbf{z}_{\textrm{r}}+\mathbf{H}_{\textrm{r}}(\bm{\eta}\odot\bm{\eta})
					-\mathbf{P}_{\textrm{r}}^{-1}\tilde{\mathbf{P}}_{\textrm{r}}\mathbf{H}_{\textrm{r}}\mathbf{B}_{\textrm{r}}\mathbf{z}_{\textrm{r}}
					-\mathbf{P}_{\textrm{r}}^{-1}\tilde{\mathbf{P}}_{\textrm{r}}\mathbf{H}_{\textrm{r}}(\bm{\eta}\odot\bm{\eta}) \nonumber \\ &\quad+\mathbf{P}_{\textrm{r}}^{-1}\tilde{\mathbf{G}}^T_\textrm{r}\mathbf{W}_{\textrm{r}}\mathbf{B}_{\textrm{r}}
\mathbf{z}_{\textrm{r}}+\mathbf{P}_{\textrm{r}}^{-1}\tilde{\mathbf{G}}^T_{\textrm{r}}\mathbf{W}_{\textrm{r}}
(\bm{\eta}\odot\bm{\eta})\nonumber \\ &\quad -\mathbf{P}_{\textrm{r}}^{-1}\tilde{\mathbf{P}}_{\textrm{r}}\mathbf{P}_{\textrm{r}}^{-1}\tilde{\mathbf{G}}^T_{\textrm{r}}\mathbf{W}_{\textrm{r}}
					\mathbf{B}_{\textrm{r}}\mathbf{z}_{\textrm{r}}
					-\mathbf{P}_{\textrm{r}}^{-1}\tilde{\mathbf{P}}_{\textrm{r}}\mathbf{P}_{\textrm{r}}^{-1}\tilde{\mathbf{G}}^T_{\textrm{r}}\mathbf{W}_{\textrm{r}}
					(\bm{\eta}\odot\bm{\eta}).
				\end{align}
				For simplicity, we can ignore the error terms in \eqref{sssb19} which are higher than the second order. As a result, using $\tilde{\mathbf{P}}_{\textrm{r}}$ in \eqref{b16}, $\Tilde{\mathbf{u}}$ is approximated as
				\begin{align}\label{b20}
					\Tilde{\mathbf{u}} &\approx\mathbf{H}_{\textrm{r}}\mathbf{B}_{\textrm{r}}\mathbf{z}_{\textrm{r}}+\mathbf{H}_{\textrm{r}}
(\bm{\eta}\odot\bm{\eta})
\nonumber \\ &\quad					-\mathbf{P}_{\textrm{r}}^{-1}\tilde{\mathbf{G}}^T_{\textrm{r}}\mathbf{W}_{\textrm{r}}\mathbf{G}_{\textrm{r}}
					\mathbf{H}_\textrm{r}\mathbf{B}_\textrm{r}\mathbf{z}_\textrm{r}
			-\mathbf{P}_\textrm{r}^{-1}\mathbf{G}^{T}_\textrm{r}\mathbf{W}_\textrm{r}\tilde{\mathbf{G}}_\textrm{r}
					\mathbf{H}_\textrm{r}\mathbf{B}_\textrm{r}\mathbf{z}_\textrm{r}
					\nonumber \\ &\quad +\mathbf{P}_{\textrm{r}}^{-1}\tilde{\mathbf{G}}^T_{\textrm{r}}\mathbf{W}_{\textrm{r}}\mathbf{B}_{\textrm{r}}\mathbf{z}_{\textrm{r}}.
				\end{align}
				As a result,  the bias of preliminary estimation  is written as
				\begin{align}\label{b21}
					\mathbb{E}[\Tilde{\mathbf{u}}]	&=\mathbb{E}[\mathbf{H}_{\textrm{r}}\mathbf{B}_{\textrm{r}}\mathbf{z}_{\textrm{r}}+\mathbf{H}_{\textrm{r}}(\bm{\eta}\odot\bm{\eta})
		\nonumber \\ &\quad			-\mathbf{P}_{\textrm{r}}^{-1}\tilde{\mathbf{G}}^T_{\textrm{r}}\mathbf{W}_{\textrm{r}}\mathbf{G}_{\textrm{r}}
					\mathbf{H}_\textrm{r}\mathbf{B}_\textrm{r}\mathbf{z}_\textrm{r}
					-\mathbf{P}_\textrm{r}^{-1}\mathbf{G}^{T}_\textrm{r}\mathbf{W}_\textrm{r}\tilde{\mathbf{G}}_\textrm{r}
					\mathbf{H}_\textrm{r}\mathbf{B}_\textrm{r}\mathbf{z}_\textrm{r}
					\nonumber \\ &\quad	+\mathbf{P}_{\textrm{r}}^{-1}\tilde{\mathbf{G}}^T_{\textrm{r}}\mathbf{W}_{\textrm{r}}\mathbf{B}_{\textrm{r}}
					\mathbf{z}_{\textrm{r}}]\nonumber \\
					&={\bf E}_1+{\bf E}_2+{\bf E}_3,
				\end{align}
				where ${\bf E}_1=\mathbb{E}[\mathbf{H}_{\textrm{r}}\mathbf{B}_{\textrm{r}}\mathbf{z}_{\textrm{r}}]$, ${\bf E}_2=\mathbb{E}[\mathbf{H}_{\textrm{r}}(\bm{\eta}\odot\bm{\eta})]$ and ${\bf E}_3=\mathbb{E}[-\mathbf{P}_{\textrm{r}}^{-1}\tilde{\mathbf{G}}^T_{\textrm{r}}\mathbf{W}_{\textrm{r}}\mathbf{G}_{\textrm{r}}
				\mathbf{H}_\textrm{r}\mathbf{B}_\textrm{r}\mathbf{z}_\textrm{r}
				-\mathbf{P}_\textrm{r}^{-1}\mathbf{G}^{T}_\textrm{r}\mathbf{W}_\textrm{r}\tilde{\mathbf{G}}_\textrm{r}
				\mathbf{H}_\textrm{r}\mathbf{B}_\textrm{r}$ $\mathbf{z}_\textrm{r}
				+\mathbf{P}_{\textrm{r}}^{-1}\tilde{\mathbf{G}}^T_{\textrm{r}}\mathbf{W}_{\textrm{r}}\mathbf{B}_{\textrm{r}}
				\mathbf{z}_{\textrm{r}}]$, the detailed derivations of which are given in Appendix \ref{appe1}.
\subsection{Bias of Refining Estimation}
				In this subsection, we aim to derive the bias of $\hat{\bf q}$ of refining estimation. Similar to the derivations of the bias of preliminary estimation, we need to derive the expression of estimation error of   ${\bm \xi}$, and the bias of ${\bm \xi}$ can be derived by taking the expectation of the estimation error of   ${\bm \xi}$.
				First, denote $\tilde{\bm \xi}$ as the estimation error of ${\bm \xi}$, e.g., $\tilde{\bm \xi} = \breve{\bm \xi}-{\bm \xi}$.
				Then, using the definition of $\breve{\bm \xi}$ in \eqref{s2_7}, $\tilde{\bm \xi}$ can be further derived as
				\begin{equation}\label{b2ss2}
					\tilde{\bm \xi}
					= ({\bf G}_1^T\hat{\bf W}_1{\bf G}_1)^{-1}{\bf G}^T_1\hat{\bf W}_1(\hat{\bf h}_1-{\bf G}_1{\bm \xi}).
				\end{equation}
				According to \eqref{s2_5}, $\tilde{\bm \xi}$ is derived as
		\begin{equation}\label{b2ss2}
					\tilde{\bm \xi} =({\bf G}_1^T\hat{\bf W}_1{\bf G}_1)^{-1}{\bf G}^T_1\hat{\bf W}_1{\bf z}_1.
				\end{equation}
				Using the derivations of \eqref{s2_6}, the definition of ${\bf B}_1$ in \eqref{s2_8} and the definition of $\tilde{\mathbf u}$ below \eqref{s2_5}, we have ${\bf z}_1={\bf B}_1\tilde{\mathbf u}+\tilde{\mathbf u}\odot\tilde{\mathbf u}$. Then,  \eqref{b2ss2} can be derived as
				\vspace{-0.1cm}
				\begin{equation}\label{b2ss3}
					\tilde{\bm \xi} = ({\bf G}_1^T\hat{\bf W}_1{\bf G}_1)^{-1}{\bf G}_1^T\hat{\bf W}_1({\mathbf B}_1\tilde{\mathbf u}+\tilde{\mathbf u}\odot\tilde{\mathbf u}).
				\end{equation}
				By defining $\hat{{\bf P}}_1={\bf G}_1^T\hat{\bf W}_1{\bf G}_1$, we have $\tilde{\bm \xi}=\hat{{\bf P}}_1^{-1}{\bf G}_1^T\hat{\bf W}_1({\mathbf B}_1\Tilde{\mathbf u}+\Tilde{\mathbf u}\odot\Tilde{\mathbf u})$. According to the definition of $\hat{\bf W}_1$ below \eqref{ab7}, $\hat{\bf W}_1$ is composed of the true value and the estimation error, which are denoted by ${\mathbf W}_1$ and $\tilde{\mathbf W}_1$, respectively. To further derive the expression of $\tilde{\bm \xi}$, the expressions of ${\mathbf W}_1$ and $\tilde{\mathbf W}_1$ should be derived, which are given by
				\begin{align}\label{b2ss5}
					\hat{\mathbf W}_1&={\mathbf W}_1+\tilde{\mathbf W}_1,
					{\mathbf W}_1=\mathbf{B}_1^{-1}{\bf P}_{\textrm{r}}\mathbf{B}_1^{-1},\nonumber \\
					\tilde{\mathbf W}_1&=\mathbf{B}_1^{-1}\tilde{\bf P}_{\textrm{r}}\mathbf{B}_1^{-1}-{\mathbf W}_1\tilde{\bf B}_1\mathbf{B}_1^{-1}-\tilde{\bf B}_1\mathbf{B}_1^{-1}{\mathbf W}_1,
				\end{align}
				where $\tilde{\bf B}_1$ is given in \eqref{b2ss4}.  The details of deriving ${\mathbf W}_1$ and $\tilde{\mathbf W}_1$ are given in Appendix \ref{ap1}.
				Similarly, based on the definition of $\hat{{\bf P}}_1$ below \eqref{b2ss3}, $\hat{{\bf P}}_1$ can be obtained as a summation of   the true value and the estimation error, which are denoted by ${\mathbf P}_1$ and $\tilde{\mathbf P}_1$, respectively.
				Hence, we have
				\begin{equation}\label{b2s5}
					\hat{\bf P}_1 = {\mathbf P}_1+\tilde{\mathbf P}_1,\quad
					{\mathbf P}_1={\bf G}_1^T{\mathbf W}_1{\bf G}_1,\quad
					\tilde{\mathbf P}_1={\bf G}_1^T\tilde{\mathbf W}_1{\bf G}_1.
				\end{equation}
				Similar to \eqref{b18}, we can derive the approximation of $\hat{\bf P}_1^{-1}$ as
				\begin{equation}\label{b2s6}
					\hat{\bf P}_1^{-1} \approx ({\bf I}-{\mathbf P}_1^{-1}\tilde{\mathbf P}_1){\mathbf P}_1^{-1}={\mathbf P}_1^{-1}-{\mathbf P}_1^{-1}\tilde{\mathbf P}_1{\mathbf P}_1^{-1}.
				\end{equation}
				Therefore, using the approximation of $\hat{\bf P}_1^{-1}$ in \eqref{b2s6} and the definition of $\hat{\bf W}_1$ in \eqref{b2ss5}, \eqref{b2ss3} can be derived as
				\begin{equation}\label{b2ss6}
					\tilde{\bm \xi}
					\approx({\mathbf P}_1^{-1}-{\mathbf P}_1^{-1}\tilde{\mathbf P}_1{\mathbf P}_1^{-1}){\bf G}_1^T\hat{\bf W}_1({\mathbf B}_1\Tilde{\mathbf u}+\Tilde{\mathbf u}\odot\Tilde{\mathbf u}).
				\end{equation}
				For notation simplicity, let ${\bf H}_1={\mathbf P}_1^{-1}{\bf G}^T_1{\mathbf W}_1$, we have
				\begin{align}\label{b2s8}
					\Tilde{\bm \xi} &\approx{\bf H}_1({\mathbf B}_1\Tilde{\mathbf u}+\Tilde{\mathbf u}\odot\Tilde{\mathbf u})+{\mathbf P}_1^{-1}{\bf G}^T_1\tilde{\mathbf W}_1({\mathbf B}_1\Tilde{\mathbf u}+\Tilde{\mathbf u}\odot\Tilde{\mathbf u})\nonumber\\
					&\quad-{\mathbf P}_1^{-1}\tilde{\mathbf P}_1{\bf H}_1({\mathbf B}_1\Tilde{\mathbf u}+\Tilde{\mathbf u}\odot\Tilde{\mathbf u})\nonumber\\
					&\quad-{\mathbf P}_1^{-1}\tilde{\mathbf P}_1{\mathbf P}_1^{-1}{\bf G}_1\tilde{\mathbf W}_1({\mathbf B}_1\Tilde{\mathbf u}+\Tilde{\mathbf u}\odot\Tilde{\mathbf u}).
				\end{align}			
				By ignoring the error terms higher than the second order and substituting $\tilde{\mathbf P}_1$ in \eqref{b2s5} into \eqref{b2s8}, we have
				\begin{equation}\label{b2s9}
					\setlength{\abovedisplayskip}{1pt}
					\setlength{\belowdisplayskip}{1pt}
					\Tilde{\bm \xi}
					\approx{\bf H}_1({\mathbf B}_1\Tilde{\mathbf u}+\Tilde{\mathbf u}\odot\Tilde{\mathbf u})+{\mathbf P}_1^{-1}{\bf G}^T_1\tilde{\mathbf W}_1{\mathbf B}_1\Tilde{\mathbf u}-{\mathbf P}_1^{-1}\tilde{\mathbf P}_1{\bf H}_1{\mathbf B}_1\Tilde{\mathbf u}.
				\end{equation}
				By assuming that ${\bf P}_2={\bf I}_{4\times4}-{\bf G}_1{\bf H}_1$, we have
				\begin{equation}\label{b2s10}
					\Tilde{\bm \xi}
					\approx{\bf H}_1({\mathbf B}_1\Tilde{\mathbf u}+\Tilde{\mathbf u}\odot\Tilde{\mathbf u})+{\mathbf P}_1^{-1}{\bf G}^T_1\tilde{\mathbf W}_1{\bf P}_2{\mathbf B}_1\Tilde{\mathbf u}.
				\end{equation}
				Finally, by taking expectation of $\Tilde{\bm \xi}$, the bias is given by			
				\vspace{-0.1cm}
				\begin{equation}\label{29}
					\setlength{\abovedisplayskip}{3pt}
					\setlength{\belowdisplayskip}{3pt}
					\mathbb{E}[\Tilde{\bm \xi}]
					={\bf H}_1{\mathbf B}_1\mathbb{E}[\Tilde{\mathbf u}]+{\bf H}_1\mathbb{E}[\Tilde{\mathbf u}\odot\Tilde{\mathbf u}]+{\mathbf P}_1^{-1}{\bf G}^T_1\mathbb{E}[\tilde{\mathbf W}_1{\bf P}_2{\mathbf B}_1\Tilde{\mathbf u}].
				\end{equation}
				To further derive the bias, let us introduce \textbf{Proposition 1} as follows.
				
				\textbf{Proposition 1.} \emph{For the  vector $\tilde{\bf a}\in\mathbb{C}^{M\times1}$, the expectation of $\tilde{\bf a}\odot\tilde{\bf a}$ can be expressed as a column vector ${\bf c}_{\textrm{a}}$ containing the diagonal elements of  ${\bm \Omega}_{\textrm{a}}$, which is the expectation of the second order moment of  $\tilde{\bf a}$.}
				
				\textbf{Proof:} Please see Appendix \ref{prop1}. $\hfill \blacksquare$

				Using \textbf{Proposition 1}, $\mathbb{E}[\Tilde{\mathbf u}\odot\Tilde{\mathbf u}]$ can be derived as ${\bf c}_{\textrm{u}}$ containing the diagonal elements of ${\bm \Omega}_{\textrm{u}}$. Furthermore, the details of deriving ${\bm \Omega}_{\textrm{u}}$ are given in Appendix A of \cite{TWu}. Therefore, the bias of $\Tilde{\bm \xi}$  can be further derived as
				\begin{equation}\label{29}
					\mathbb{E}[\Tilde{\bm \xi}]={\bf H}_1({\mathbf B}_1\mathbb{E}[\Tilde{\mathbf u}]+{\bf c}_{\textrm{u}})+{\mathbf P}_1^{-1}{\bf G}^T_1\mathbb{E}[\tilde{\mathbf W}_1{\bf P}_2{\mathbf B}_1\Tilde{\mathbf u}],
				\end{equation}
				where $\mathbb{E}[\Tilde{\mathbf u}]$ is given in \eqref{b21}, and the derivations of $\mathbb{E}[\tilde{\mathbf W}_1{\bf P}_2{\mathbf B}_1\Tilde{\mathbf u}]$ are given in Appendix \ref{ap2}.
\begin{figure*}[t]
    \centering
    \begin{minipage}[b]{0.24\linewidth}
        \centering
        \includegraphics[width=\textwidth]{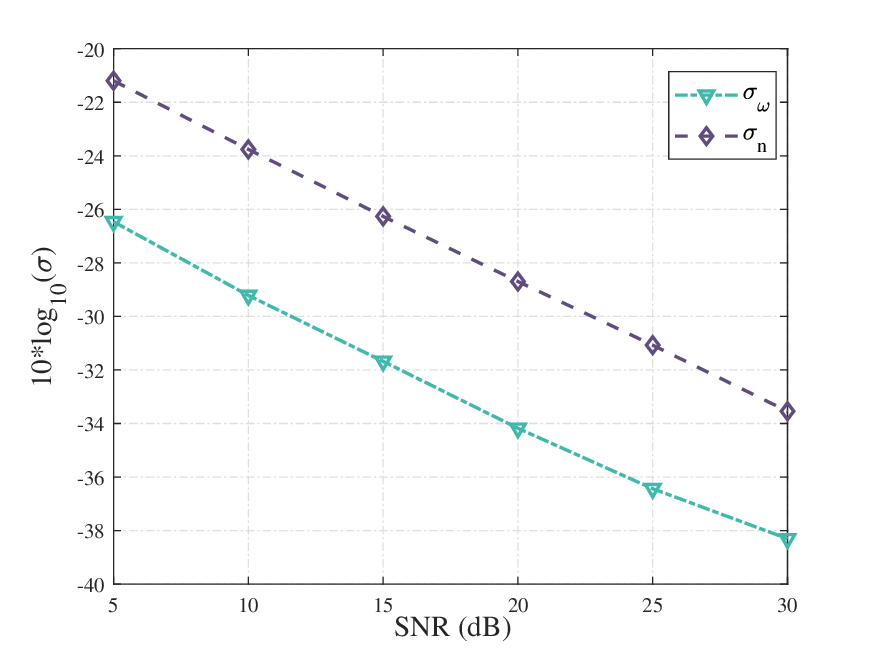}
        \caption{Standard deviation of $n_{i}$ and $\omega_{i}$ versus SNR (dB). \vspace{5ex}}
        \label{std_SNR}
    \end{minipage}
    \hfill
    \begin{minipage}[b]{0.24\linewidth}
        \centering
        \includegraphics[width=\textwidth]{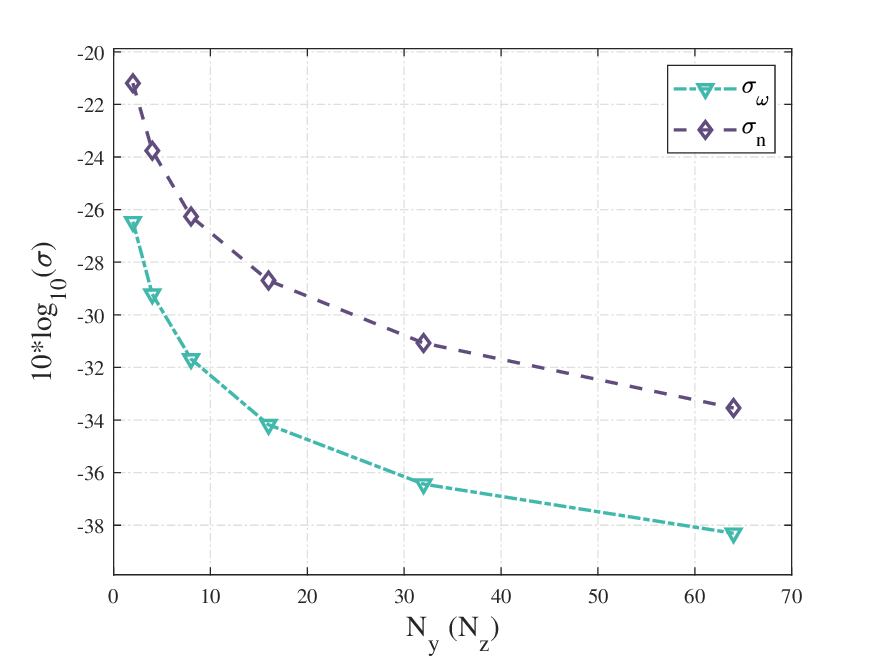}
        \caption{Standard deviation of $n_{i}$ and $\omega_{i}$ versus $\textrm{N}_\textrm{y}$$(\textrm{N}_\textrm{z})$.\vspace{5ex}}
        \label{std_num}
    \end{minipage}
    \hfill
    \begin{minipage}[b]{0.24\linewidth}
        \centering
        \includegraphics[width=\textwidth]{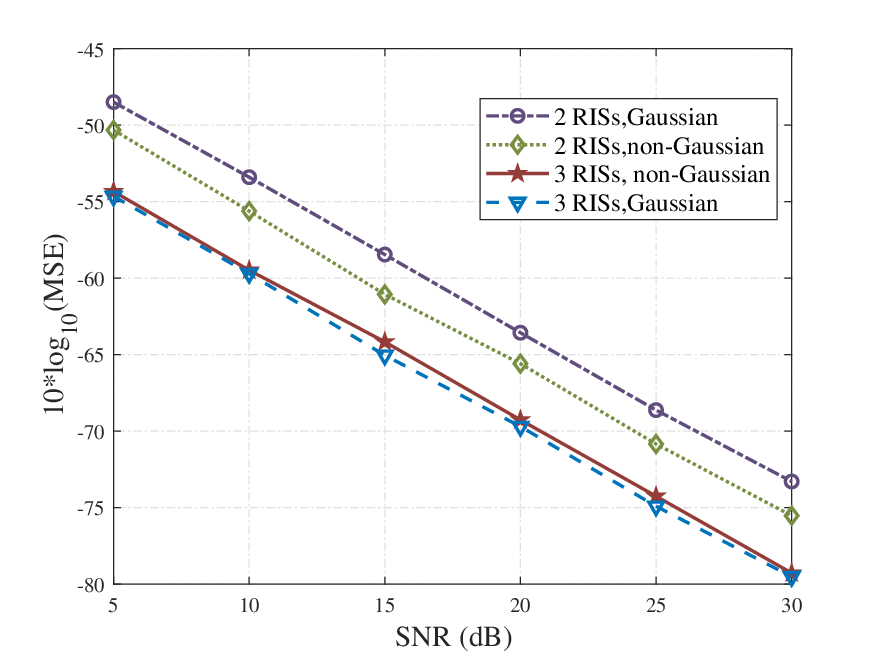}
        \caption{The MSE comparison of the proposed algorithm in the cases of Gaussian and non-Gaussian versus SNR (dB).}
        \label{SNR_Gau}
    \end{minipage}
    \hfill
    \begin{minipage}[b]{0.24\linewidth}
        \centering
        \includegraphics[width=\textwidth]{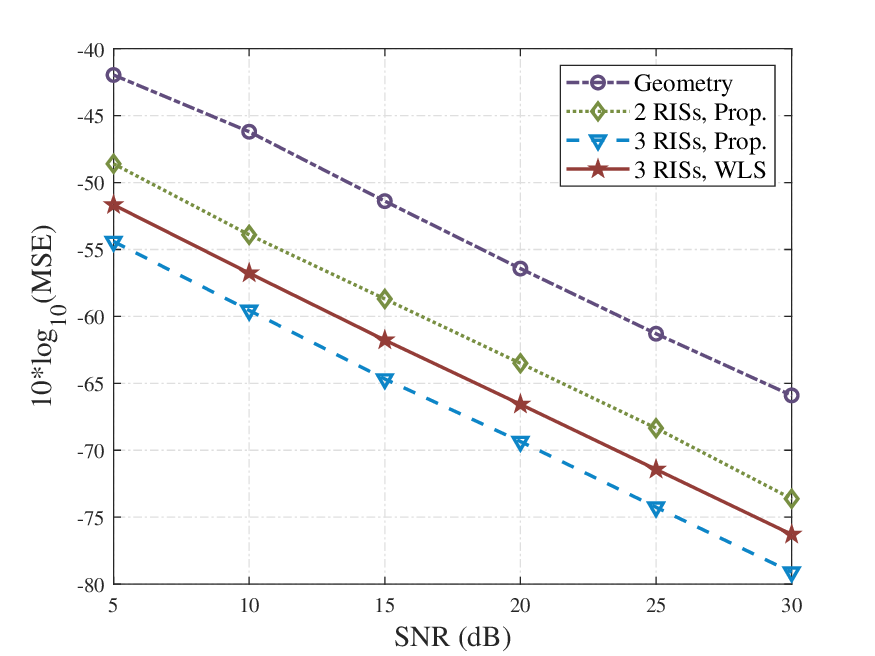}
        \caption{The MSE comparison of the proposed algorithm with other algorithms versus SNR (dB).}
        \label{SNR_Alg}
    \end{minipage}
\end{figure*}

								Then, let us define the estimation error of MU's position as $\tilde{\bf q}=\hat{\bf q}-{\bf q}$. Then using the definition of ${\bm \xi}$ in \eqref{s2_3}, $\breve{\bm \xi}$ can be derived as		
				\begin{equation}\label{bswe1}
					\setlength{\abovedisplayskip}{3pt}
					\setlength{\belowdisplayskip}{3pt}
					\breve{\bm \xi} =\left(\hat{{\bf q}}-{\bf p}\right)\odot\left(\hat{{\bf q}}-{\bf p}\right)
					=\tilde{\bf q}\odot\tilde{\bf q}+2\tilde{\bf q}\odot({\bf q}-{\bf p})+{\bm \xi}.
				\end{equation}
				By utilizing $\Tilde{\bm \xi}=\breve{\bm \xi}-{\bm \xi}$, we have  $\Tilde{\bm \xi}-\tilde{\bf q}\odot\tilde{\bf q}=2\tilde{\bf q}\odot({\bf q}-{\bf p})$. Then, by assuming that ${\bf B}_q\tilde{\bf q}=2\tilde{\bf q}\odot({\bf q}-{\bf p})$, where ${\bf B}_q= 2 \textrm{diag}\{({\bf q}-{\bf p})\}$, we have the expression of $\tilde{\bf q}$ given by		
				\begin{equation}\label{bs1}\setlength{\abovedisplayskip}{3pt}
					\setlength{\belowdisplayskip}{3pt}		
					\tilde{\bf q}={\bf B}_q^{-1}(\Tilde{\bm \xi}-\tilde{\bf q}\odot\tilde{\bf q}).
				\end{equation}
				Then, we can derive the bias of refining estimation as		
				\vspace{-0.1cm}
				\begin{equation}\label{bs2}
					\setlength{\abovedisplayskip}{3pt}
					\setlength{\belowdisplayskip}{3pt}
					\mathbb{E}[\tilde{\bf q}]=\mathbb{E}[{\bf B}_q^{-1}(\Tilde{\bm \xi}-\tilde{\bf q}\odot\tilde{\bf q})].
				\end{equation}
				Using \textbf{Proposition 1}, we have $\mathbb{E}[\tilde{\bf q}\odot\tilde{\bf q}]={\bf c}_q$, where ${\bf c}_q$ is a column vector formed by the diagonal elements of ${\bm \Omega}_q$, and the details of deriving ${\bm \Omega}_q$ can be found in Appendix \ref{ap3}. Therefore, \eqref{bs2} can be further derived as
				\begin{equation}\label{bs21}
					\mathbb{E}[\tilde{\bf q}]={\bf B}_q^{-1}(\mathbb{E}[\Tilde{\bm \xi}]-{\bf c}_q),
				\end{equation}
				where $\mathbb{E}[\Tilde{\bm \xi}]$ is given in \eqref{29}.

			\section{Simulation Results}\label{simulation}
				This section presents simulation results to evaluate the performance of the proposed localization algorithm aided by multiple RISs. Moreover, the MU, the BS, and the RISs are assumed to be placed in a 3D  area. The location of the BS is  ${\bf p}=[10,12,12]^T$, while the locations of three RISs are ${\bf s}_1=[2,20,2]^T$, ${\bf s}_2=[-12,-16,58]^T$ and ${\bf s}_3=[-10,-10,50]^T$, respectively.  The phase shift matrix of the RIS is set to a unit matrix. The following results are obtained by averaging over 10,000 random estimation error realizations. The localization accuracy is assessed in terms of the  MSE, and the bias. The TDoA and AoA estimation error are assumed to follow the Gaussian distribution. In the figures showing the simulation results, we shall use the log scale for the error level to indicate the wide range of the levels tested.  	

Fig.\ref{std_SNR} shows  the standard deviations of $n_{i}$ and $\omega_{i}$ as the functions of  SNR (dB). It is observed that the standard deviations of $n_{i}$ and $\omega_{i}$ decrease with the RIS size, which means that increasing the number of elements could improve the estimation accuracy. It is observed that the standard deviations  decrease with SNR increasing, which validates that  improved communication setting is capable of enhancing angle estimation accuracy. Besides, it is observed that the standard deviation $\sigma_{\omega}$ is smaller than $\sigma_{n}$. This implies that the estimation accuracy for $ {\phi}_{MR,i}$ is superior to that of $ {\theta}_{MR,i}$. This phenomenon is hypothesized to be due to the sequential estimation process in 2D-DFT. Initially, $ {\phi}_{MR,i}$ is estimated, followed by $ {\theta}_{MR,i}$. The sequential nature of this process leads to an accumulation of errors, resulting in a greater estimation error for $ {\theta}_{MR,i}$.

Fig.\ref{std_num}  illustrates the standard deviations of $n_{i}$ and $\omega_{i}$ as the functions of the RIS size, denoted as ${N}_{y}$ and ${N}_{z}$. Furthermore, it is shown that once ${N}_{y}$ and ${N}_{z}$ exceeds $32$, increasing the panel size of the RIS does not significantly improve the accuracy of angle estimation. This indicates that there is a diminishing return on the benefits of increasing the number of RIS elements. In practical communication scenarios, it may be sufficient to maintain a certain number of RIS elements without continually expanding them.

Fig. \ref{SNR_Gau} clearly demonstrates the MSE comparisons  in cases of Gaussian and non-Gaussian AoA estimation errors. As shown in the figure, the MSE decreases  with SNR as expected. Furthermore, it is evaluated that the localization accuracy of the systems employing $3$ RISs is much higher than that of the systems employing $2$ RISs, which effectively demonstrates the importance of the number of the RISs.

 More importantly, it can be observed from Fig. \ref{SNR_Gau} that the MSE obtained from the localization algorithm using practical non-Gaussian errors does not align with that using Gaussian errors. This suggests that in designing practical localization algorithms, it is preferable to base them on actual non-Gaussian errors. It is important to note that algorithms designed using actual non-Gaussian errors are more adept at reflecting practical error characteristics. This contrasts with algorithms based on Gaussian errors, which typically rely on idealized assumptions of normal distribution, potentially leading to inaccuracies in practical applications. The inconsistency observed in the  MSE results between these two approaches underscores the gap between theoretical models and actual scenarios. This further substantiates the advantage of designing an algorithm based on non-Gaussian errors, as it can more accurately reflect and adapt to the error characteristics encountered in practical applications.

 As shown in Fig.\ref{SNR_Alg}, when comparing the performance in terms of MSE, the geometry algorithm \cite{Henk}, which relies on geometric relationships for positioning estimates, shows approximately 10 times higher MSE compared to the WLS and the proposed algorithms. On the other hand, the proposed algorithm consistently outperforms the WLS algorithm, achieving approximately 5 times lower MSE. These results strongly demonstrate the superior performance of our proposed  algorithm within the framework. These findings reinforce the effectiveness and superiority of our proposed  algorithm within the framework.
\begin{figure*}[t]
    \centering
    \begin{minipage}{0.32\linewidth}
        \centering
        \includegraphics[width=\textwidth]{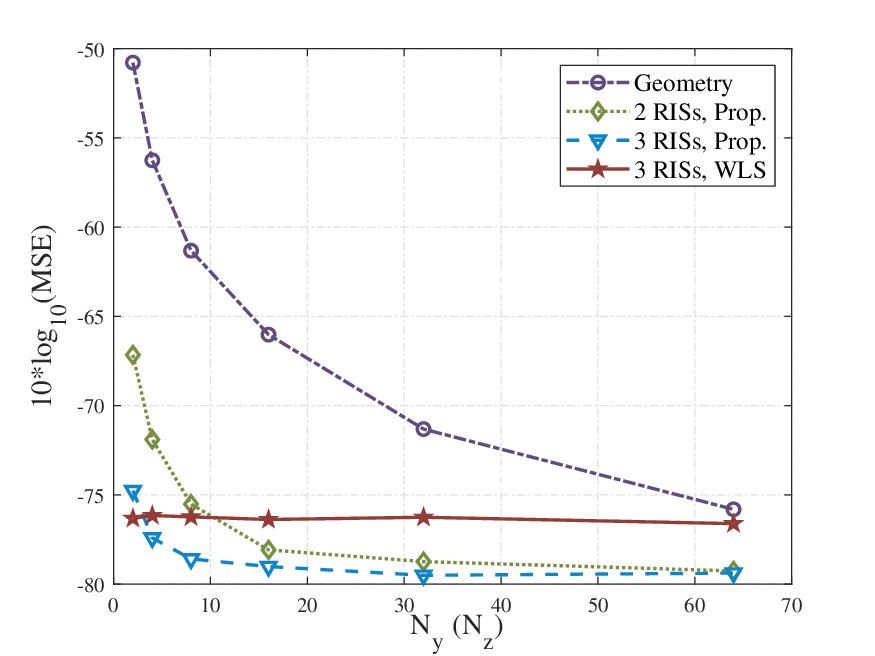}
        \caption{The MSE comparison of the proposed algorithm with other algorithms versus RIS size $\textrm{N}_\textrm{y}$ $(\textrm{N}_\textrm{z})$.}
        \label{num_Alg}
    \end{minipage}
    \hfill
    \begin{minipage}{0.32\linewidth}
        \centering
        \includegraphics[width=\textwidth]{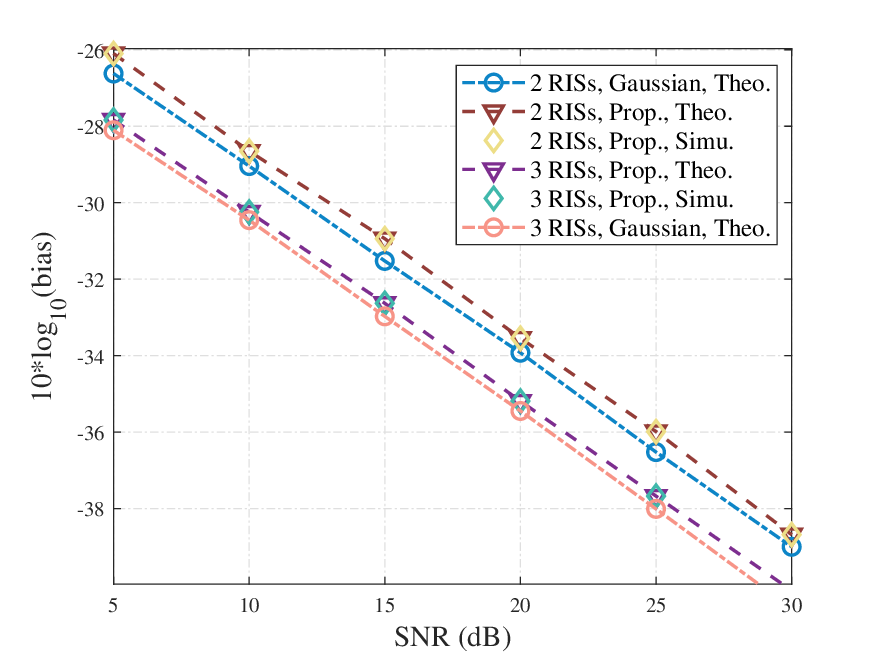}
        \caption{The bias comparison of the proposed algorithm in the cases of Gaussian and non-Gaussian versus SNR (dB).}
        \label{SNR_Bias}
    \end{minipage}
    \hfill
    \begin{minipage}{0.32\linewidth}
        \centering
        \includegraphics[width=\textwidth]{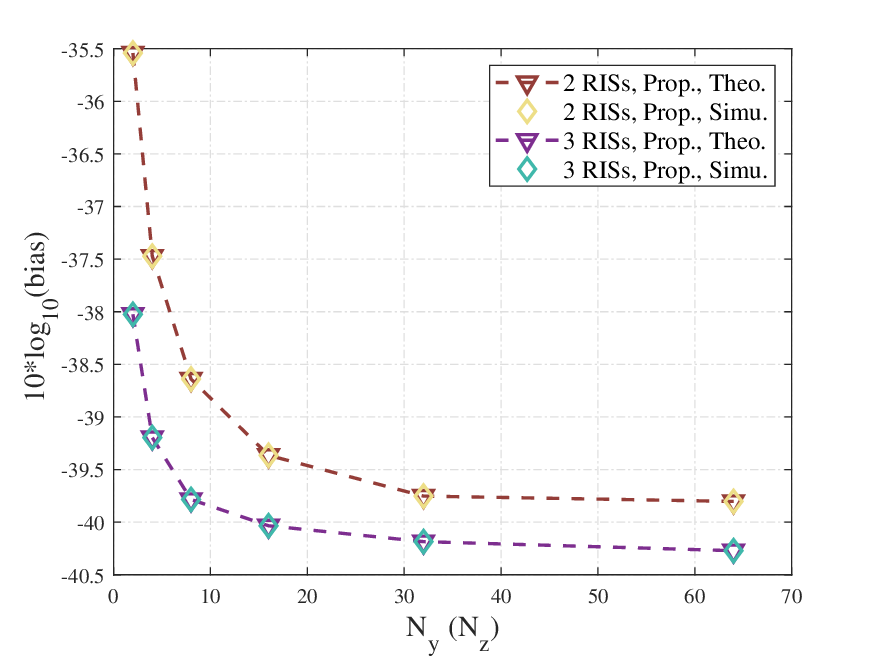}
        \caption{The bias comparison of the proposed algorithm versus RIS size $\textrm{N}_\textrm{y}$ $(\textrm{N}_\textrm{z})$.}
        \label{num_Bias}
    \end{minipage}
\end{figure*}

Based on the angle estimation expressions presented in \eqref{1em18}, it is evident that the estimation accuracy is influenced by the number of RIS elements. Accordingly, simulation results are provided in Fig.\ref{num_Alg}, comparing the MSE across various algorithms. The results, as depicted in Fig.\ref{num_Alg}, reveal that our proposed algorithm demonstrates superior performance over other algorithms. Additionally,  it observed that beyond a certain threshold, specifically when ${N}_{y}$ and ${N}_{z}$ exceeds $16$, the increase in the number of RIS elements does not significantly enhance the precision of the positioning system. Furthermore, the use of $3$ RISs  does not notably improve the positioning accuracy compared to a configuration with $2$ RISs. This suggests a point of diminishing returns in terms of positioning accuracy relative to the number of the RISs and the number of RIS elements  employed. This phenomenon may be due to the fact that the optimization of the system is already approaching its limits at lower RIS numbers, or that other factors (e.g., signal interference, hardware limitations, etc.) are starting to dominate at higher RIS numbers.

Fig.\ref{SNR_Bias} and Fig.\ref{num_Bias}  compare the bias of the proposed localization algorithm aided by 2 and 3 RIS, respectively, thereby validating the accuracy of our theoretical analysis on the algorithm's bias. Specifically, Fig.\ref{SNR_Bias} focuses on contrasting the bias in cases of Gaussian and non-Gaussian AoA estimation errors. The simulation results, as illustrated in Fig.\ref{SNR_Bias}, align more closely with the theoretical outcomes for non-Gaussian AoA estimation errors, rather than Gaussian ones. This alignment suggests that for a more accurate depiction of localization performance in real-world wireless systems, it is preferable to consider practical non-Gaussian AoA estimation errors in performance analyses. Additionally, these findings underscore the advantage of our bias analysis approach, which proves to be more tractable in practical scenarios.

				\section{Conclusion}\label{Con}
				In this paper, we investigated  the algorithm design and analysis for RIS-aided localization IoT system aided by multiple RISs, tackling the non-Gaussian AEE. We designed the mWLS algorithm to derive the closed-form expression of the position of the MU.  Finally, we investigated the bias analysis to evaluate the performance of the proposed algorithm. Simulation results confirmed the accuracy of the derived results and demonstrated the localization performance of the proposed algorithm.

				\begin{appendices}
					
					\section{}\label{appe1}
					By utilizing \eqref{26}, recall that $\mathbf{z}_{\textrm{r}}=[\mathbf{n}^T,\bm{\omega}^T,\bm{\nu}^T]^T$, $\mathbf{B}_\textrm{r}\mathbf{z}_\textrm{r}$  can be expanded as $(\mathbf{B}_{{\textrm{r},n}}\mathbf{n}+\mathbf{B}_{{\textrm{r},}\omega}\bm{\omega}+\mathbf{B}_{{\textrm{r},}\nu}\bm{\nu})$. Then, we have
\begin{align}
						{\bf E}_1 &= \mathbb{E}[\mathbf{H}_{\textrm{r}}\mathbf{B}_{\textrm{r}}\mathbf{z}_{\textrm{r}}]
						=\mathbf{H}_{\textrm{r}}\mathbb{E}[\mathbf{B}_{{\textrm{r},n}}\mathbf{n}+
						\mathbf{B}_{{\textrm{r},}\omega}\bm{\omega}+\mathbf{B}_{{\textrm{r},}\nu}\bm{\nu}]\nonumber\\
						&=\mathbf{H}_{\textrm{r}}\mathbf{B}_{{\textrm{r},n}}\mathbb{E}[\mathbf{n}]+
						\mathbf{H}_{\textrm{r}}\mathbf{B}_{{\textrm{r},}\omega}\mathbb{E}[\bm{\omega}],\nonumber
\end{align}
where $\mathbb{E}[\mathbf{n}]$ and $\mathbb{E}[\bm{\omega}]$ denote the expectation of $\mathbf{n}$ and $\bm{\omega}$, respectively.

					For ${\bf E}_2$, we have
					\begin{align}
						{\bf E}_2 &= \mathbb{E}[\mathbf{H}_{\textrm{r}}(\bm{\eta}\odot\bm{\eta})]\nonumber\\
					&=\frac{1}{2}\mathbf{H}_{\textrm{r}}\mathbb{E}[[\underbrace{0,\cdots,0}_{2M},\nu^2_1,\cdots,\nu^2_M]^T]=\frac{1}{2}\mathbf{H}_{\textrm{r}}\mathbf{q}_u,\nonumber
					\end{align}
					where
					\begin{equation}
						\setlength{\abovedisplayskip}{3pt}
						\setlength{\belowdisplayskip}{3pt}
						\mathbf{q}_u=\frac{1}{2}[\underbrace{0,\cdots,0}_{2M},\sigma^2_{\nu_1},\cdots,\sigma^2_{\nu_M}]^T.\nonumber
					\end{equation}
					
					Then, ${\bf E}_3$ can be derived as
					\begin{equation}
						\setlength{\abovedisplayskip}{1pt}
						\setlength{\belowdisplayskip}{1pt}
					\begin{aligned}
						{\bf E}_3
						&= \mathbb{E}[\mathbf{P}_{\textrm{r}}^{-1}\tilde{\mathbf{G}}^T_{\textrm{r}}\mathbf{W}_{\textrm{r}}\mathbf{B}_{\textrm{r}}
						\mathbf{z}_{\textrm{r}}]\nonumber\\
					&\quad-\mathbb{E}[\mathbf{P}_{\textrm{r}}^{-1}\tilde{\mathbf{G}}^T_{\textrm{r}}\mathbf{W}_{\textrm{r}}
						\mathbf{G}_{\textrm{r}}
						\mathbf{H}_\textrm{r}\mathbf{B}_\textrm{r}\mathbf{z}_\textrm{r}]
						-\mathbb{E}[\mathbf{P}_\textrm{r}^{-1}\mathbf{G}^{T}_\textrm{r}\mathbf{W}_\textrm{r}\tilde{\mathbf{G}}_\textrm{r}
						\mathbf{H}_\textrm{r}\mathbf{B}_\textrm{r}\mathbf{z}_\textrm{r}
						]\nonumber\\
						&={\bf E}_{31}-{\bf E}_{32}-{\bf E}_{33},\nonumber
					\end{aligned}\end{equation}
					where the detailed derivations of ${\bf E}_{31}$, ${\bf E}_{32}$ and ${\bf E}_{33}$ are given as follows.
					
					For ${\bf E}_{31}$, as we have $\tilde{\bf G}_{\textrm{r}}$ in \eqref{bb18} and  $\mathbf{B}_\textrm{r}\mathbf{z}_\textrm{r}=(\mathbf{B}_{{\textrm{r},n}}\mathbf{n}+\mathbf{B}_{{\textrm{r},}\omega}
					\bm{\omega}+\mathbf{B}_{{\textrm{r},}\nu}\bm{\nu})$,   ${\bf E}_{31}$ can be derived as
					\begin{equation}
						\begin{aligned}
						{\bf E}_{31}
						&=\mathbb{E}[\mathbf{P}_{\textrm{r}}^{-1}(\bar{\mathbf{G}}_{n}^T\odot [(\mathbf{n1}_{1\times4})^T,(\mathbf{n1}_{1\times4})^T,(\mathbf{n1}_{1\times4})^T])\nonumber\\
					&\quad\quad\quad\mathbf{W}_{\textrm{r}}
						\mathbf{B}_{{\textrm{r},n}}\textrm{diag}({\bf n}){\bf 1}_{M\times1}]\nonumber \\
						&\quad+\mathbb{E}[\mathbf{P}_{\textrm{r}}^{-1}(\bar{\mathbf{G}}^T_{\omega}\odot [(\bm{\omega1}_{1\times4})^T,(\bm{\omega1}_{1\times4})^T,(\bm{\omega1}_{1\times4})^T])\nonumber\\
					&\quad\quad\quad\mathbf{W}_{\textrm{r}}
						\mathbf{B}_{{\textrm{r},\omega}}\textrm{diag}({\bm \omega}){\bf 1}_{M\times1}]\nonumber \\
						&\quad+\mathbb{E}[\mathbf{P}_{\textrm{r}}^{-1}(\bar{\mathbf{G}}^T_{\nu}\odot [(\bm{\nu1}_{1\times4})^T,(\bm{\nu1}_{1\times4})^T,(\bm{\nu1}_{1\times4})^T])\nonumber\\
					&\quad\quad\quad\mathbf{W}_{\textrm{r}}
						\mathbf{B}_{{\textrm{r},\nu}}\textrm{diag}({\bm \nu}){\bf 1}_{M\times1}].
						\nonumber		
					\end{aligned}
					\end{equation}
					To further derive the expression of ${\bf E}_{31}$, we introduce \textbf{Proposition 2} and \textbf{Proposition 3} as follows.

					\textbf{Proposition 2.} \emph{For vectors} ${\bm \alpha} \in \mathbb{C}^{M\times 1} $, ${\bm \beta} \in \mathbb{C}^{N\times 1}$ \emph{and matrix } ${\bf G} \in \mathbb{C}^{M\times N}$, \emph{and corresponding diagonal matrices} $\textrm{diag}({\bm \alpha})$ \emph{and} $\textrm{diag}({\bm \beta})$ \emph{with these vectors as their main diagonals, we 	have}	$({\bm \alpha}{\bm \beta}^T)\odot {\bf G}=\textrm{diag}({\bm \alpha}){\bf G}\textrm{diag}({\bm \beta})$.		
					
					\textbf{Proof:}   Please see reference \cite{horn_johnson_1985}.$\hfill \blacksquare$
					
					\textbf{Proposition 3.} \emph{For a vector ${\bf x} \in \mathbb{C}^{M\times1}$ with its covariance matrix given by ${\bf Q}_x$, we have}		
					\begin{align}
						&\mathbb{E}[({\bf A}\odot[({\bf x1}_{1\times N})^T,({\bf x1}_{1\times N})^T,({\bf x1}_{1\times N})^T]){\bf B}\textrm{diag}({\bf x})]\nonumber\\
					&=
						{\bf A}({\bf B}\odot[{\bf Q}_x,{\bf Q}_x,{\bf Q}_x]^T),\nonumber
					\end{align}
					\emph{where ${\bf A}\in\mathbb{C}^{N\times 3M}$ and ${\bf B}\in\mathbb{C}^{3M\times M}$.}
					
					\textbf{Proof:}  Please see reference \cite{TWu}.

					\section{}\label{ap1}
					First, based on the definition of $\hat{\mathbf W}_1$ below \eqref{ab7} and the definition of $\hat{\bf P}_{\textrm{r}}$ above \eqref{b14}, we have $\hat{\mathbf W}_1=\hat{\mathbf{B}}_1^{-1}\hat{\bf P}_{\textrm{r}}\hat{\mathbf{B}}_1^{-1}$. Then, according to the definition of ${\mathbf B}_1$ in \eqref{s2_8} and $\hat{\mathbf B}_1$ in \eqref{ab7}, we have
					\begin{equation}\label{b2ss4}
						\hat{\bf B}_1 = \mathbf{B}_1+\tilde{\bf B}_1,
						\tilde{\bf B}_1 = 2 \textrm{diag} \{\tilde{\bf u}\}.
					\end{equation}
					If we use $\hat{\bf B}_1=\mathbf{B}_1+\tilde{\bf B}_1$ to  derive $\hat{\mathbf W}_1$, the inverse matrix  $\hat{\bf B}_1^{-1}=(\mathbf{B}_1+\tilde{\bf B}_1)^{-1}$ would be complex and challenging. Fortunately, using the Newmann expansion \cite{2012Bias}, we can derive the approximation of $\hat{\bf B}_1^{-1}
					\approx\mathbf{B}_1^{-1}-\mathbf{B}_1^{-1}\tilde{\bf B}_1\mathbf{B}_1^{-1}$. As a result, $\hat{\bf W}_1$ is given by
					\begin{equation}\label{b2s3}
						\hat{\bf W}_1
						=(\mathbf{B}_1^{-1}-\mathbf{B}_1^{-1}\tilde{\bf B}_1\mathbf{B}_1^{-1})({\bf P}_{\textrm{r}}+\tilde{\bf P}_{\textrm{r}})(\mathbf{B}_1^{-1}-\mathbf{B}_1^{-1}\tilde{\bf B}_1\mathbf{B}_1^{-1}).
					\end{equation}
					By neglecting the error terms that are higher than the first order, $\hat{\bf W}_1$ can be approximated as
					\begin{align}
						\label{b2s3we}
						\hat{\bf W}_1
						&\approx\mathbf{B}_1^{-1}{\bf P}_{\textrm{r}}\mathbf{B}_1^{-1}+\mathbf{B}_1^{-1}\tilde{{\bf P}}_{\textrm{r}}\mathbf{B}_1^{-1}-\mathbf{B}_1^{-1}{\bf P}_{\textrm{r}}\mathbf{B}_1^{-1}\tilde{\bf B}_1\mathbf{B}_1^{-1}\nonumber\\
					&\quad-\mathbf{B}_1^{-1}\tilde{\bf B}_1\mathbf{B}_1^{-1}{\bf P}_{\textrm{r}}\mathbf{B}_1^{-1}
						\nonumber\\
					&={\mathbf W}_1+\tilde{\mathbf W}_1,
					\end{align}
					where ${\bf W}_1$ and $\tilde{\mathbf W}_1$ are defined as			
					\begin{align}\label{b2s4}
						{\mathbf W}_1&=\mathbf{B}_1^{-1}{\bf P}_{\textrm{r}}\mathbf{B}_1^{-1},\nonumber\\
						\tilde{\mathbf W}_1&=\mathbf{B}_1^{-1}\tilde{\bf P}_{\textrm{r}}\mathbf{B}_1^{-1}-{\mathbf W}_1\tilde{\bf B}_1\mathbf{B}_1^{-1}-\tilde{\bf B}_1\mathbf{B}_1^{-1}{\mathbf W}_1.
					\end{align}
\section{}\label{prop1}
					Before proving \textbf{Proposition 1}, let us introduce \textbf{Proposition 4}  as follows.
					
					\textbf{Proposition 4.} \emph{The Hadamard product of two vectors } ${\bf a} \in \mathbb{C}^{N \times 1}$ \emph{and} ${\bf b} \in \mathbb{C}^{N \times 1}$, \emph{ is the same as matrix multiplication of one vector by the corresponding diagonal matrix of the other vector:}
					\begin{equation}
												\setlength{\abovedisplayskip}{1pt}
						\setlength{\belowdisplayskip}{1pt}
						{\bf a}\odot {\bf b}=\textrm{diag}({\bf a}){\bf b}.\nonumber
					\end{equation}

					\textbf{Proof:}   please see reference \cite{horn_johnson_1985}. $\hfill \blacksquare$

					Let $\mathbb{E}[\tilde{\bf a}\odot\tilde{\bf a}]$ denote the expectation of $\tilde{\bf a}\odot\tilde{\bf a}$, using \textbf{Proposition 4},  we have $\mathbb{E}[\tilde{\bf a}\odot\tilde{\bf a}]
					=\mathbb{E}[\textrm{diag}(\tilde{\bf a})\tilde{\bf a}]=\mathbb{E}[\textrm{diag}(\tilde{\bf a})\textrm{diag}(\tilde{\bf a}){\bf 1}_{M\times1}]$.
					Using \textbf{Proposition 2},  $\mathbb{E}[\tilde{\bf a}\odot\tilde{\bf a}]$ can be rewritten as
					\begin{align}
												\setlength{\abovedisplayskip}{1pt}
						\setlength{\belowdisplayskip}{1pt}
						&\mathbb{E}[\tilde{\bf a}\odot\tilde{\bf a}]\nonumber\\
					&=\mathbb{E}[((\tilde{\bf a}\tilde{\bf a}^T)\odot{\bf I}_{M \times M}){\bf 1}_{M\times1}]\nonumber\\
					&
						=({\bm \Omega}_{\textrm{a}}\odot{\bf I}_{M \times M}){\bf 1}_{M\times1}
						={\bf c}_{\textrm{a}},\nonumber
					\end{align}
					where ${\bf c}_{\textrm{a}}$ denotes the  vector containing the diagonal elements of  ${\bm \Omega}_{\textrm{a}}$. Here, we  complete the proof of \textbf{Proposition 1}.

					\section{}\label{ap2}
					Using the definition of $\tilde{\mathbf W}_1$ in \eqref{b2s4},  $\mathbb{E}[\tilde{\mathbf W}_1{\bf P}_2{\mathbf B}_1\Tilde{\mathbf u}]$ can be derived as
				\begin{align}
					\setlength{\abovedisplayskip}{1pt}
					\setlength{\belowdisplayskip}{1pt}
					\begin{aligned}
						&\mathbb{E}[\tilde{\mathbf W}_1{\bf P}_2{\mathbf B}_1\Tilde{\mathbf u}]\nonumber\\
&=\mathbb{E}[(\mathbf{B}_1^{-1}\tilde{\bf P}_{\textrm{r}}\mathbf{B}_1^{-1}-{\mathbf W}_1\tilde{\bf B}_1\mathbf{B}_1^{-1}-\tilde{\bf B}_1\mathbf{B}_1^{-1}{\mathbf W}_1){\bf P}_2{\mathbf B}_1\Tilde{\mathbf u}]\nonumber\\
						&=\mathbb{E}[\mathbf{B}_1^{-1}\tilde{\bf P}_{\textrm{r}}\mathbf{B}_1^{-1}{\bf P}_2{\mathbf B}_1\Tilde{\mathbf u}]-\mathbb{E}[{\mathbf W}_1\tilde{\bf B}_1\mathbf{B}_1^{-1}{\bf P}_2{\mathbf B}_1\Tilde{\mathbf u}]\nonumber\\
					&\quad-\mathbb{E}[\tilde{\bf B}_1\mathbf{B}_1^{-1}{\mathbf W}_1{\bf P}_2{\mathbf B}_1\Tilde{\mathbf u}].\nonumber
					\end{aligned}\end{align}
					Let ${\mathbf a}_1=\mathbb{E}[\mathbf{B}_1^{-1}\tilde{\bf P}_{\textrm{r}}\mathbf{B}_1^{-1}{\bf P}_2{\mathbf B}_1\Tilde{\mathbf u}]$, ${\mathbf a}_2=-\mathbb{E}[{\mathbf W}_1\tilde{\bf B}_1\mathbf{B}_1^{-1}{\bf P}_2{\mathbf B}_1\Tilde{\mathbf u}]$ and ${\mathbf a}_3=-\mathbb{E}[\tilde{\bf B}_1\mathbf{B}_1^{-1}{\mathbf W}_1{\bf P}_2{\mathbf B}_1\Tilde{\mathbf u}]$, we have $\mathbb{E}[\tilde{\mathbf W}_1{\bf P}_2{\mathbf B}_1\Tilde{\mathbf u}]={\mathbf a}_1+{\mathbf a}_2+{\mathbf a}_3$, while ${\mathbf a}_1$, ${\mathbf a}_2$, ${\mathbf a}_3$ are given as follows.
					
					Using the definition of $\tilde{\bf P}_{\textrm{r}}$ below \eqref{b16}, ${\mathbf a}_1$ can be derived as
					\begin{align}
						\setlength{\abovedisplayskip}{1pt}
						\setlength{\belowdisplayskip}{1pt}
	\begin{aligned}
						{\mathbf a}_1&=\mathbb{E}[\mathbf{B}_1^{-1}\tilde{\bf P}_{\textrm{r}}\mathbf{B}_1^{-1}{\bf P}_2{\mathbf B}_1\Tilde{\mathbf u}]\nonumber\\
					&
						=\mathbb{E}[\mathbf{B}_1^{-1}(\tilde {\bf G}^T_{\textrm{r}}{\bf W}_{\textrm{r}}{\bf G}_{\textrm{r}}
						+{\bf G}^{T}_{\textrm{r}}{\bf W}_{\textrm{r}}\tilde{\bf G}_{\textrm{r}})\mathbf{B}_1^{-1}{\bf P}_2{\mathbf B}_1\Tilde{\mathbf u}].\nonumber
					\end{aligned}\end{align}
					Using the definition of $\Tilde{\mathbf u}$ in \eqref{b20} and ignoring the error terms that are higher than the second order, ${\mathbf a}_1$ can be approximated as $\mathbb{E}[\mathbf{B}_1^{-1}(\tilde {\bf G}^T_{\textrm{r}}{\bf W}_{\textrm{r}}{\bf G}_{\textrm{r}}
					+{\bf G}^{T}_{\textrm{r}}{\bf W}_{\textrm{r}}\tilde{\bf G}_{\textrm{r}})\mathbf{B}_1^{-1}{\bf P}_2{\mathbf B}_1{\bf H}_{\textrm{r}}{\bf B}_{\textrm{r}}{\bf z}_{\textrm{r}}]$. For notation simplicity, we assume that $\mathbf{B}_1^{-1}{\bf P}_2{\mathbf B}_1{\bf H}_{\textrm{r}}={\bf P}_3$. Then, by applying   \textbf{Proposition 2} and  \textbf{Proposition 3}, ${\mathbf a}_1$ can be derived as
			\begin{align}
				\setlength{\abovedisplayskip}{1pt}
				\setlength{\belowdisplayskip}{1pt}
					\begin{aligned}
						{\mathbf a}_1&=\mathbf{B}_1^{-1}(\bar{{\bf G}}^T_n(({\bf W}_{\textrm{r}}{\bf G}_{\textrm{r}}{\bf P}_3{\bf B}_{\textrm{r},n})\odot\bar{{\bf Q}}_n){\bf 1}_{M\times1}\nonumber\\
						&\quad+\bar{{\bf G}}^T_\omega(({\bf W}_{\textrm{r}}{\bf G}_{\textrm{r}}{\bf P}_3{\bf B}_{\textrm{r},\omega})\odot\bar{{\bf Q}}_\omega){\bf 1}_{M\times1}\nonumber\\
						&\quad+\bar{{\bf G}}^T_\nu(({\bf W}_{\textrm{r}}{\bf G}_{\textrm{r}}{\bf P}_3{\bf B}_{\textrm{r},\nu})\odot\bar{{\bf Q}}_\nu){\bf 1}_{M\times1}\nonumber\\
						&\quad
						+{\bf G}^{T}_{\textrm{r}}(({\bf W}_{\textrm{r}}\bar{{\bf G}}_n{\bf P}_3{\bf B}_{\textrm{r},n})\odot\bar{{\bf Q}}_n){\bf 1}_{M\times1}\nonumber\\
						&\quad+{\bf G}^{T}_{\textrm{r}}(({\bf W}_{\textrm{r}}\bar{{\bf G}}_\omega{\bf P}_3{\bf B}_{\textrm{r},\omega})\odot\bar{{\bf Q}}_\omega){\bf 1}_{M\times1}
						\nonumber\\
						&\quad+{\bf G}^{T}_{\textrm{r}}(({\bf W}_{\textrm{r}}\bar{{\bf G}}_\nu{\bf P}_3{\bf B}_{\textrm{r},\nu})\odot\bar{{\bf Q}}_\nu){\bf 1}_{M\times1}).\nonumber
					\end{aligned}\end{align}
						
					Moreover, for ${\mathbf a}_2$, using the definition of $\tilde{\bf B}_1$ in \eqref{b2ss4},  we have
					\begin{align}
						\setlength{\abovedisplayskip}{1pt}
						\setlength{\belowdisplayskip}{1pt}
						\begin{aligned}
						{\mathbf a}_2&=-\mathbb{E}[{\mathbf W}_1\mathbf{B}_1^{-1}\tilde{\bf B}_1{\bf P}_2{\mathbf B}_1\Tilde{\mathbf u}]\nonumber\\
						&=-2{\mathbf W}_1\mathbf{B}_1^{-1}\mathbb{E}[\textrm{diag}(\tilde{\bf u}){\bf P}_2{\mathbf B}_1\textrm{diag}(\tilde{\bf u}){\bf 1}_{4\times1}].\nonumber
					\end{aligned}\end{align}
					Using \textbf{Proposition 2}, we have
					\begin{align}
						\setlength{\abovedisplayskip}{1pt}
						\setlength{\belowdisplayskip}{1pt}
					\begin{aligned}
						{\mathbf a}_2	&=-2{\mathbf W}_1\mathbf{B}_1^{-1}\mathbb{E}[(\tilde{\bf u}\tilde{\bf u}^T)\odot({\bf P}_2{\bf B}_1)]{\bf 1}_{4\times1}\nonumber\\
						&=-2{\mathbf W}_1\mathbf{B}_1^{-1}({\bf \Omega}_{\textrm{u}}\odot({\bf P}_2{\bf B}_1)){\bf 1}_{4\times1}.\nonumber
					\end{aligned}\end{align}
					As ${\bf \Omega_{\textrm{u}}}$ is a diagonal matrix, $	{\mathbf a}_2$ can be further derived as
					\begin{equation}
						\begin{aligned}
						{\mathbf a}_2&=-2{\mathbf W}_1\mathbf{B}_1^{-1}\textrm{diag}({\bf \Omega}_{\textrm{u}}({\bf P}_2{\bf B}_1)){\bf 1}_{4\times1}=-2{\mathbf W}_1\mathbf{B}_1^{-1}{\bf p}_\textrm{4},\nonumber
					\end{aligned}\end{equation}
					where ${\bf p}_\textrm{4}$  represents the column vector formed by the diagonal elements of ${\bf P}_4={\bf P}_2{\mathbf B}_1{\bf \Omega}_\textrm{u}$.
					
					Similar to the derivations of ${\bf a}_2$, ${\mathbf a}_3$ can be derived as
					\begin{equation}
						\setlength{\abovedisplayskip}{1pt}
						\setlength{\belowdisplayskip}{1pt}
						\begin{aligned}
						{\mathbf a}_3=-\mathbb{E}[\mathbf{B}_1^{-1}\tilde{\bf B}_1{\mathbf W}_1{\bf P}_2{\mathbf B}_1\tilde{\mathbf u}]=-2\mathbf{B}_1^{-1}{\bf p}_\textrm{4,W1},\nonumber
					\end{aligned}\end{equation}
					where ${\bf p}_\textrm{4,W1}$ represents the column vector formed by the diagonal elements of  ${\mathbf W}_1{\bf P}_4$.

					\section{}\label{ap3}
					First, let us derive the covariance matrix of ${\bm \xi}$, which is denoted by ${\bf \Omega}_{\xi}$. Using the definition of $\tilde{\bm \xi}$ in \eqref{b2ss2},
					its covariance matrix ${\bf \Omega}_{\xi}$ is given by
				\begin{align}
					\setlength{\abovedisplayskip}{1pt}
					\setlength{\belowdisplayskip}{1pt}
                    \begin{aligned}
						{\bf \Omega}_{\xi}&=\mathbb{E}[\tilde{\bm \xi}\tilde{\bm \xi}^T]\nonumber\\
						&\approx{\bf G}_1^T\hat{\bf W}_1{\bf G}_1)^{-1}{\bf G}^T_1\hat{\bf W}_1\hat{\bf W}_1^{-1}\hat{\bf W}_1^T{\bf G}_1
						(({\bf G}_1^T\hat{\bf W}_1{\bf G}_1)^{-1})^T
						\nonumber\\
						&=({\bf G}_1^T\hat{\bf W}_1{\bf G}_1)^{-1},\nonumber
					\end{aligned}\end{align}
					where $\mathbb{E}({\bf z}_1{\bf z}_1^T)\approx\hat{\bf \Omega}_1=\hat{\bf W}_1^{-1}$ is given below \eqref{ab7}.
					
					Furthermore, by neglecting the second order term in \eqref{bs1}, the $\tilde{\bf q}$ can be approximated as ${\bf B}_q^{-1}\Tilde{\bm \xi}$, therefore, the covariance matrix of $\hat{\bf q}$ can be derived as
				\begin{equation}
					\begin{aligned}
						{\bf \Omega}_{q}&=\mathbb{E}[\tilde{\bf q}\tilde{\bf q}^T]= {\bf B}_q^{-1}({\bf G}_1^T\hat{\bf W}_1{\bf G}_1)^{-1}{\bf B}_q^{-1}.\nonumber
					\end{aligned}\end{equation}
				\end{appendices}

				\bibliographystyle{IEEEtran}
				\bibliography{myre}

			\end{document}